\documentclass[twocolumn,showpacs,prb]{revtex4}%
\usepackage{amsmath}
\usepackage{graphicx}
\usepackage{dcolumn}
\usepackage{bm}%
\usepackage{amsfonts}%
\usepackage{amssymb}
\providecommand{\U}[1]{\protect\rule{.1in}{.1in}}
\begin{document}
\title{Microscopic analysis of the superconducting quantum critical point: Finite
temperature crossovers in transport near a pair-breaking quantum phase
transition }
\author{Nayana Shah}
\affiliation{Department of Physics, University of Illinois at Urbana-Champaign, 1110 W.
Green St, Urbana, IL 61801, USA}
\author{Andrei Lopatin}
\affiliation{Materials Science Division, Argonne National Laboratory, Argonne, Illinois
60439, USA. }

\begin{abstract}
A microscopic analysis of the superconducting quantum critical point realized
via a pair-breaking quantum phase transition is presented. Finite temperature
crossovers are derived for the electrical conductivity, which is a key probe
of superconducting fluctuations. By using the diagrammatic formalism for
disordered systems, we are able to incorporate the interplay between
fluctuating Cooper pairs and electrons, that is outside the scope of a
time-dependent Ginzburg Landau or effective bosonic action formalism. It is
essential to go beyond the standard approximation in order to capture the zero
temperature correction which results purely from the (dynamic) quantum
fluctuations and dictates the behavior of the conductivity in an entire low
temperature quantum regime. All dynamic contributions are of the same order
and conspire to add up to a negative total, thereby inhibiting the
conductivity as a result of superconducting fluctuations. On the contrary, the
classical and the intermediate regimes are dominated by the positive bosonic
channel. Our theory is applicable in one, two and three dimensions and is
relevant for experiments on superconducting nanowires, doubly-connected
cylinders, thin films and bulk in the presence of magnetic impurities,
magnetic field or other pair-breakers. A window of non-monotonic behavior is
predicted to exist as either the temperature or the pair-breaking parameter is swept.

\end{abstract}

\maketitle

\section{Introduction}

The study of quantum phase transitions\cite{SondhiGCS1997,Sachdev} has
consistently been one of the frontier fields in condensed matter physics, for
the last decades. The steady growth in the possibility of simultaneously
accessing lower temperatures and higher values of tuning parameters such as
pressure and magnetic field, together with the possibility of operating old
and new experimental techniques under such extreme conditions, has provided
the necessary thrust. Discovery of new materials and physical systems has been
another key factor. Not only has it maintained the freshness and novelty of
the field but also helped to identify the universal features. Complex
materials usually go hand in hand with complex phase diagrams with many
competing or coexisting orders, given the multitude of energy scales that they
have. Although this accounts for a rich body of physics, it also makes it
harder to unmask the universal features and to systematically explore the
neighborhood of the quantum phase transitions in the system.

Technological advances in relatively recent years have made it possible to
make precisely designed and controllable bulk, mesoscopic and nano systems,
consisting of cold atoms and quantum dots, for example. These have not only
opened new avenues for studying quantum phase transitions but also served as
model systems where theoretical predictions can be verified with the help of
tunable parameters. This in turn can act as a starting point for understanding
complex materials which are not as easy to control.

By now there is a large body of theoretical work devoted to studying quantum
phase transitions in a whole range of systems ranging from heavy fermion
compounds, high temperature superconductors, manganites and organic materials,
to quantum dot and cold atomic systems. Transitions between and out of
different correlated states of matter ranging from magnetic, superconducting,
and charge-ordered states to more exotic, fractional and topological states,
have been the subject of research. In spite of active interest, many questions
still remain unanswered.

To-date superconductivity remains one of the most striking examples of
emergent many-body states and quantum phase transitions involving a
superconducting state are ubiquitous in a whole variety of materials. In many
a cases, it appears as one of the multiple phases in a complex phase diagram,
with its mechanism not always clearly understood. However there is another set
of materials consisting of single elements or simple compounds with relatively
simple phase diagrams which too display superconductivity.

The BCS theory that was put forward exactly fifty years ago was able to
explain the basic mechanism of superconductivity which in turn was discovered
for the first time almost another fifty years earlier. This theory is one of
the crowning glories of condensed matter physics and a prototype of an ideal
many body theory, which is successful in explaining the experimentally
observed thermodynamic and electrodynamic behavior in a large class of
materials which is commonly labelled as conventional or
BCS\ superconductors\cite{Tinkham,Gennes}.

The point of view we want to take in this paper is to study the physics of
quantum phase transitions in such a BCS\ superconductor for which the theory
is well-understood and experimental properties well-characterized. The theory
could then be cleanly tested by taking a conventional superconductor with a
simple phase diagram. Such a study would exemplify the physics of a
superconducting quantum phase transition which is not only of fundamental and
technological interest but also a prototype for quantum phase transitions in
other correlated systems.

The key characteristic of the BCS\ theory is the pairing of electrons with
their time-reversed partners to form a condensate of Cooper pairs that
superconducts. As the temperature is increased more and more quasi-particles
are formed by exciting electrons out of the condensate by breaking the pairs,
until superconductivity is destroyed at a mean-field transition temperature.
The experimental transition temperature is well-defined and the transition is
mean-field like in most conventional bulk superconductors. Superconducting
fluctuations however do exist even beyond the transition temperature and the
fluctuations become stronger as the dimensionality of the sample is reduced.
The effect of fluctuating Cooper pairs on different physical properties such
as diamagnetism, specific heat, and conductivity, was actively studied --both
theoretically and experimentally-- about a decade after the BCS\ theory was
put forward; see the review on fluctuation effects in Ref.
[\onlinecite{SkocpolT1975}]
which was already written three decades ago. The interest in studying
fluctuation effects saw a revival after the discovery of high-temperature
superconductivity and corresponding theoretical results can be found in a more
recent review article\cite{LarkinV}.

What is the way to start from a superconducting state at the absolute zero of
the temperature and destroy it at a finite value of some tuning parameter, via
a second order phase transition? One has to think of a way of breaking the
pairs without the help of thermal fluctuations. It turns out that all one has
to do is to turn on a perturbation that breaks the time-reversal symmetry.
Pair-breaking perturbations resulting in a suppression of the transition
temperature have been well-understood and it has been known for a long time
that superconductivity is destroyed even at zero temperature, once the
pair-breaking parameter reaches a critical strength. We recognize such
pair-breaking quantum phase transitions (with dynamic critical exponent $z=2$)
out of a superconducting state as an important class of quantum phase
transitions and make it the subject of our study.

In spite of the long history of superconducting fluctuation effects near the
classical transition, there has been no systematic theoretical analysis of
such effects in the vicinity of a pair-breaking quantum phase transition. In
this paper we have used the finite temperature Matsubara diagrammatic method
to evaluate the fluctuation corrections to the electrical conductivity which
is one of the key physical quantities capable of diagnosing and show-casing
the role of superconducting fluctuations. We have incorporated the (dynamic)
quantum fluctuations by going beyond the standard approximations and mapped
out the fluctuation regimes near the transition. Such a study is timely given
the progress in fabricating ultra-narrow superconducting nanowires and
doubly-connected cylinders in addition to thin-film samples. Not only will it
allow identifying the universal features near superconducting quantum phase
transitions in complex materials, but also play a crucial role in enhancing
our understanding of mesoscopic superconductivity. The microscopic approach we
use allows us to treat the fluctuation corrections involving the interaction
between the electrons and the fluctuating Cooper pairs and identify the
regimes in which they dominate. Our analysis can then serve as a guideline for
the construction and validation of an effective bosonic theory or a
time-dependent Ginzburg Landau formalism which does not incorporate the
electrons. Such a bosonic theory can then be used for further analysis of the
quantum critical point especially in one and two dimensions, given that the
upper critical dimension in our case is two.

The outline of the paper is as follows: In Sec. II we lay the foundation for
the remainder of the paper, by clearly defining the meaning of a pair-breaking
perturbation and introducing the class of pair-breaking quantum phase
transitions that are of interest to us. We present the framework based on
Usadel equations as a systematic and general method for obtaining the
expression for the pair-breaking parameter in a given situation and illustrate
it by considering some examples which are simple yet relevant for our analysis.

In Sec. III we present the calculation of the fluctuation corrections to the
normal state conductivity in the vicinity of the pair-breaking phase
transition. We introduce the key building blocks of the temperature
diagrammatic perturbation theory that we need, show how they are modified in
the presence of a pair-breaking perturbation, and present their limiting
forms. A fairly detailed account of the actual evaluation of diagrams
corresponding to the fluctuation corrections is then given, focusing on the
careful considerations required in order to incorporate the quantum (dynamic)
fluctuations correctly. The results of this section are applicable near the
entire phase transition line, starting from the classical finite temperature
transition in the absence of pair-breaking perturbation to the quantum phase
transition driven by tuning the pair-breaking parameter at zero temperature.

The evaluation of these general expressions in the vicinity of the quantum
phase transition is done in Sec. IV. Once the dominant corrections are
identified, we present the different fluctuation regimes that come out of our
analysis. There are three regimes --quantum, intermediate and classical-- and
the behavior of the conductivity depends on the path of approach to the
quantum phase transition. We demonstrate how these results vary based on the
effective dimensionality of the problem and provide predictions of our theory
that should be applicable to experiments on nanowires or doubly-connected
cylinders, thin films, and bulk systems. The result we obtain by evaluating
the \textquotedblleft Aslamazov-Larkin\textquotedblright\ correction in the
vicinity of the classical transition is given as a benchmark to compare with
the well-established results in the literature and also with its behavior near
the quantum phase transition.

In the final section we place our work in a bigger picture by discussing
related theoretical and experimental work. Different theoretical approaches
for the same problem as well as slightly different physical configurations
studied using a formalism similar to ours, are both included. An attempt to
interpret all the experiments on superconducting quantum phase transition in
thin films would clearly be outside the scope of this paper. We have hence
focussed mainly on those that have analyzed the data in terms of quantum
corrections, including the fluctuation corrections. Relevant experiments on
superconducting nanowires and doubly-connected cylinders are relatively scant
so far. However we discuss the current status to support our belief that given
the technological advances in the recent years, the predictions of our theory
should not only be accessible but also important from the point of view of
using ultra narrow wires in superconducting electronic circuits.

\section{Pair-breaking parameter ($\alpha$)}

\subsection{Definition and physical meaning}

A\ pair-breaking perturbation is any perturbation that breaks the
time-reversal degeneracy of a superconducting paired state. Anderson's
theorem\cite{Anderson1959} asserts that in the absence of such a perturbation,
the superconducting critical temperature $T_{c}$ and the BCS\ density of
states remains the same even after alloying the superconductor with
impurities. However if the impurities are magnetic, Abrikosov and
Gorkov\cite{AbrikosovG1961} found that the $T_{c}$ is suppressed and the
density of states is modified as well (giving a gapless regime). They
parametrized the strength of the pair-breaking perturbation by a
\textit{pair-breaking parameter }$\alpha$\textit{\ }and obtained
\begin{equation}
\ln\left(  \frac{T_{c}}{T_{c0}}\right)  =\psi\left(  \frac{1}{2}\right)
-\psi\left(  \frac{1}{2}+\frac{\alpha}{2\pi T_{c}}\right) \label{Tc_vs_H}%
\end{equation}
where $\psi$ is the digamma function and $T_{c0}\equiv T_{c}(\alpha=0)$. The
parameter $\alpha$ was shown to be inverse of the spin-flip scattering time
which is proportional to the density of magnetic impurities.

Following this classic work, it was recognized that the above equation for
$T_{c}$ suppression (as well as gapless superconductivity)\ can be transcribed
for a whole class of pair-breaking perturbations for which the transition to
the normal state in the presence of $\alpha$ is of second order, once the
appropriate $\alpha$ is used for each case\cite{Gennes,Tinkham}. What is
essential is the presence of a rapid scattering mechanism that modulates over
time the pair-breaking perturbation seen by a given Cooper pair of electrons,
assuring an \textit{ergodic} behavior of electrons. Then $\alpha\ $can be
interpreted as the depairing energy (energy splitting) of a pair of
time-reversed electrons, averaged over a time interval $\tau_{K}$ it takes for
their\ relative phase to be randomized by the perturbation; one thus has
$2\alpha\equiv\hbar/\tau_{K} $.

This generalization of the concept of pair-breaking offers the possibility of
defining a pair-breaking parameter not only for bulk systems but also for
mesoscopic and non-homogeneous systems and for situations in which multiple
pair-breaking mechanisms might be operative. However, the derivation of
$\alpha$ might not be straightforward in such cases. Below we will describe
the Usadel equation formalism as a general method to derive $\alpha$ for a
given configuration.

\subsection{Derivation using Usadel equations}

In this paper we are interested in focussing on dirty superconductors for
which the Usadel equation formalism \cite{Usadel1970,Kopnin2001} is
well-suited. Writing the Heisenberg equation of motion starting from the
BCS\ Hamiltonian, one gets the microscopic Gorkov equations for the normal and
anomalous Green functions. Using the fact that the characteristic length scale
for the normal state is smaller than the length scale for the superconducting
order parameter variation, one can make the quasiclassical approximation
\begin{equation}
\frac{\hbar/p_{F}}{\xi_{0}}\sim\frac{\Delta}{E_{F}}\ll1
\end{equation}
(which is quite accurate for most classic low-temperature superconductors and
less so for high temperature superconductors) to exclude the fast oscillations
of the Green functions associated with variations of the relative coordinate
on a scale $\hbar/p_{F}$ and rewrite the Gorkov equations in terms of the
quasiclassical Green functions with only a slow dependence on the
centre-of-mass coordinate varying on the scale $\xi_{0}$. The Eilenberger
equations so derived can be further simplified in the dirty limit
\begin{equation}
\frac{l}{\xi_{0}}\sim\frac{\tau}{\hbar/T_{c}}\ll1,
\end{equation}
(where $l$ is the mean free path and $\tau$ is the mean free impurity
scattering time) in which the strong scattering by impurities produces
averaging over momentum directions, to obtain the Usadel
equations\cite{Usadel1970}%
\begin{align}
-iD[g(\mathbf{\nabla}-\frac{2ie}{c}\mathbf{A)}^{2}f-f\mathbf{\nabla}^{2}g]  &
=2\Delta g-2i\omega_{n}f,\\
-iD[g(\mathbf{\nabla}+\frac{2ie}{c}\mathbf{A)}^{2}f^{\dagger}-f^{\dagger
}\mathbf{\nabla}^{2}g]  & =2\Delta^{\ast}g-2i\omega_{n}f^{\dagger}%
\end{align}
where $D=v_{F}^{2}\tau/3$ is the three-dimensional diffusion constant. Note
that $g$ and $f$, the quasiclassical Green functions averaged over momentum
directions, get expressed as functionals of the fluctuating order parameter
field $\Delta(x,\tau)$. Below we outline the precise recipe for deriving the
pair-breaking parameter, starting from these equations.

To identify the pair-breaking parameter it is enough to consider the Usadel
equations for the $f$ function to the lowest order in $\Delta$, so as to
obtain%
\begin{equation}
-iD\,g_{0}(\mathbf{\nabla}-\frac{2ie}{c}\mathbf{A})^{2}f=-2i\omega f+2\Delta
g_{0}\label{Usadeleq}%
\end{equation}
with $g_{0}=\operatorname{sign}\omega$. One could interpret the depairing
parameter as the lowest eigenvalue of the operator (in the transverse
direction)%
\begin{equation}
-{\frac{D}{2}}\,(\nabla_{\perp}-\frac{2ie}{c}\mathbf{A})^{2}\,f=\alpha
f\label{eigenvalue}%
\end{equation}
obtained by appropriately choosing the gauge and the boundary conditions (to
ensure absence of current perpendicular to a wire or film surface, for
example). Alternatively one can solve Eq. (\ref{Usadeleq}) for $f$ and read
out the expression for $\alpha$ from it. In the remaining part of this
subsection we illustrate the procedure for some examples of interest. The
formalism is general enough and can be appropriately adapted to complicated
situations involving the simultaneous occurrence of multiple pair-breaking perturbations.

Although the pair-breaking parameter for the classic example of magnetic
impurities can be derived within the Eilenberger-Usadel formalism, we refer
the reader to the original paper\cite{AbrikosovG1961} and proceed to discuss
some other specific examples.

\subsubsection{Thin Film}

Consider a thin film with thickness $s$ smaller than the superconducting
coherence length $\xi$ such that the superconducting fluctuations are
effectively two-dimensional ($d=2$). Let us focus on the orbital pair-breaking
effect of a magnetic field $H\ $applied parallel to the film. We consider $s$
to be smaller than the penetration depth $\lambda$ so that one can assume the
field to be uniform inside the sample. If the film is placed in the $xy-$plane
and $H$ is parallel to the $y-$axis, we choose the gauge such that $A_{x}=Hz,$
$A_{y}=A_{z}=0$, where $z$ is measured from the mid-plane of the film. Now Eq.
(\ref{Usadeleq}) takes the form
\begin{equation}
-i\frac{D}{2}\left[  \frac{\partial^{2}}{\partial z^{2}}+\left(  \frac{2ie}%
{c}Hz\right)  ^{2}\right]  f=-i\left\vert \omega\right\vert f+\Delta\
\end{equation}
which on integrating over the transverse direction (i.e. over $z$) leads to
\begin{equation}
iD\left(  \frac{eH}{c}\right)  ^{2}\frac{s^{3}}{6}f=-i\left\vert
\omega\right\vert fs+s\Delta
\end{equation}
on using the appropriate boundary conditions that require the transverse
derivative of $f$ to vanish on the surface of the film. We thus have%
\begin{equation}
f=\frac{-i\Delta}{\left\vert \omega\right\vert +\alpha}%
\end{equation}
with
\begin{equation}
\alpha=\frac{D}{6}\left(  \frac{eHs}{c}\right)  ^{2}\label{alpha_film}%
\end{equation}
identified as the pair-breaking parameter.

\subsubsection{Nanowire}

Consider a wire of radius $R$ such that the diameter is smaller than $\xi$ and
$\lambda$ and the effective dimensionality of the problem is $d=1$ as far as
the superconducting fluctuations are concerned. To obtain the expression for
the pair-breaking parameter coming from the orbital effect of a magnetic field
$H\ $applied parallel to the wire, we use the cylindrical coordinates and
choose the gauge such that $A_{\phi}=H\rho/2,$ $A_{\rho}=A_{z}=0.$ For the
sake of illustration, this time we start with Eq. (\ref{eigenvalue}) which
takes the form
\begin{equation}
{\frac{D}{2}}\,\left[  -\frac{1}{\rho}\frac{\partial}{\partial\rho}\left(
\rho\frac{\partial}{\partial\rho}\right)  +\left(  \frac{eH\rho}{c}\right)
^{2}\right]  \,f=\alpha f
\end{equation}
and on integrating in the transverse direction (i.e. over $\rho$) gives%
\begin{equation}
{\frac{D}{2}}\,\,f%
{\displaystyle\int\limits_{0}^{R}}
\rho d\rho\left(  \frac{eH\rho}{c}\right)  ^{2}=\alpha f%
{\displaystyle\int\limits_{0}^{R}}
\rho d\rho
\end{equation}
using the boundary conditions that require $\partial f/\partial\rho$ to vanish
at $\rho=R.$ The expression for the pair-breaking parameter
\begin{equation}
\alpha=\frac{D}{4}\left(  \frac{eHR}{c}\right)  ^{2}\label{alphawire}%
\end{equation}
is then immediately evident.

If one considers a magnetic field applied perpendicular to the wire, the
pair-breaking parameter is given by
\begin{equation}
\alpha=\frac{D}{2}\left(  \frac{eHR}{c}\right)  ^{2}\label{alphawireperp}%
\end{equation}
instead and the calculation follows on the lines similar to that for a field
applied parallel to a film.

\subsubsection{Doubly-connected cylinder}

Consider a doubly-connected (hollow) cylinder with inner radius $r_{1}$ and
outer radius $r_{2}$ such that the wall of the cylinder is thinner than $\xi$
and $\lambda$ and the effective dimensionality is $d=1$ as for the case of a
nanowire. Due to the single-valuedness of $\Delta$ and $f$ their $\phi
-$dependence is given by $e^{in\phi}$ where $n$ is an arbitrary integer.
Choosing the cylindrical gauge
\begin{equation}
\mathbf{A=}\frac{1}{2}\mathbf{h}\times\mathbf{r=}\frac{1}{2}\widehat
{\mathbf{\phi}}h\rho
\end{equation}
as we did also for the nanowire, Eq. (\ref{eigenvalue}) becomes
\begin{equation}
\ {\frac{D}{2}}\left[  -\frac{1}{\rho}\frac{d}{d\rho}\left(  \rho\frac
{d}{d\rho}\right)  +\ \left(  \frac{m}{\rho}-\frac{eh\rho}{c}\right)
^{2}\right]  f=\alpha f\
\end{equation}
Integrating over the radial direction (from $r_{1}$ to $r_{2}$), we obtain the
expression
\begin{equation}
\alpha=D\left[  {\frac{{eH}}{{4c}}}\left[  -4n+{\frac{{eH}}{{c}}}(r_{1}%
^{2}+r_{2}^{2})\right]  +n^{2}{\frac{{\ln(r_{2}/r_{1})}}{{r_{2}^{2}-r_{1}^{2}%
}}}\right]  ,\label{alpha_cylinder}%
\end{equation}
where $n$ is an arbitrary integer (note that for $r_{1}=0$ we correctly
recover the result for the nanowire). For a thin cylinder ($r_{1}\approx
r_{2}\approx r$) the pair-breaking parameter reduces to
\begin{equation}
\alpha=(D/2r^{2})(\Phi/\Phi_{0}-n)^{2}\label{alpha_cyl}%
\end{equation}
where $\Phi$ is the flux enclosed by the cylinder, thereby rendering the
classic Little-Park oscillations\cite{Tinkham} of $T_{c}$ as can be seen from
Eq. (\ref{Tc_vs_H}). Interestingly, for a cylinder with small enough radius,
$r<r_{c}=\sqrt{D\gamma/4\pi T_{co}},$ it is possible to push the $T_{c}$ down
to zero at magnetic fields corresponding to half-integer fluxes $\Phi=\Phi
_{0}(1/2+n).$

\subsection{Pair-breaking phase transition}

The boundary between the superconducting and normal state in the $\alpha-T$
plane is given by%

\begin{equation}
\ln\left(  \frac{T}{T_{c0}}\right)  =\psi\left(  \frac{1}{2}\right)
-\psi\left(  \frac{1}{2}+\frac{\alpha}{2\pi T}\right) \label{transline}%
\end{equation}
At a given pair-breaking strength $\alpha$, superconductivity is destroyed at
$T=T_{c}(\alpha)$ and at a given temperature $T$, at $\alpha=\alpha_{c}(T),$
obtained by solving Eq. (\ref{transline}) for $T$ and for $\alpha, $
respectively. In the absence of a pair-breaking perturbation ($\alpha=0$) the
system undergoes the classical transition at $T_{c}(0)\equiv$ $T_{c0}$. In the
neighborhood of this classical transition, for $\alpha\ll T_{c0}$ we can
define the quantity
\begin{equation}
\delta_{T}(\alpha,T)\equiv\frac{T-T_{c}(\alpha)}{T_{c}(\alpha)}\label{deltaT}%
\end{equation}
that measures the relative distance from the critical temperature
$T_{c}(\alpha)$.

On the other hand, if the pair-breaking effect is sufficiently strong,
superconductivity is destroyed even at $T=0$ thereby yielding a second order
quantum phase transition. The critical value of $\alpha$ at zero temperature,
\[
\alpha_{c0}\equiv\alpha_{c}(T=0)=\frac{\pi T_{c0}}{2\gamma}%
\]
can be obtained by using the first term in the asymptotic form of the digamma
function for large arguments ($\psi(z)=\ln z$). Here $\ln\gamma\approx0.577$
is the Euler constant and one can immediately see that $2\alpha_{c0}%
=1.76T_{c}=\Delta_{0},$ the BCS\ gap at zero temperature. Expanding the
RHS\ of Eq. (\ref{transline}) to next order in small $T$, one finds that the
transition curve close to $\alpha_{c0}$ is given by
\begin{equation}
\alpha_{c}(T\ll\alpha_{c0})=\alpha_{c0}-\frac{\pi^{2}T^{2}}{6\alpha_{c0}%
}\label{alphaexp}%
\end{equation}
In the vicinity of the the pair-breaking quantum phase transition we will
define the quantity
\begin{equation}
\delta_{\alpha}(\alpha,T)\equiv\frac{\alpha-\alpha_{c}(T)}{\alpha_{c}%
(T)}\label{deltaalpha}%
\end{equation}
which can be interpreted as the relative distance from the critical
pair-breaking strength $\alpha_{c}(T)$ at a given $T\ll\alpha_{c0}$.

In the following section we will be interested in evaluating the fluctuation
corrections to the normal state conductivity in the vicinity of a
pair-breaking transition. The effective quasiclassical approach based on the
Usadel equations that we have discussed in the previous subsection is quite
accurate in the dirty limit and has no applicability restrictions in terms of
the temperature range. To evaluate the fluctuation corrections, one could
imagine using the functional formalism based on this approach. However we have
chosen to use the standard diagrammatic method instead.%

\begin{figure}
[ptb]
\begin{center}
\includegraphics[
height=2.6212in,
width=3.3201in
]%
{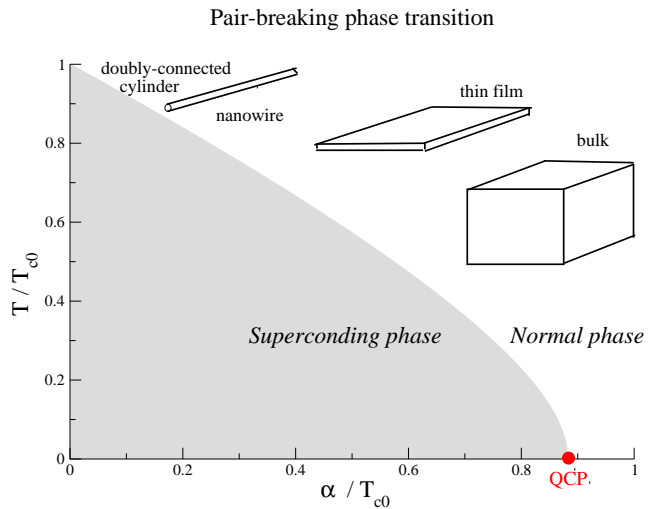}%
\caption{Phase diagram showing the pair-breaking transition from
superconducting to normal state with the boundary given by Eq.
(\ref{transline}). A superconducting quantum critical point is seen to be
realized when the pair-breaking parameter reaches a critical strength
$\alpha_{c0}/T_{c0}=0.889$. Different systems of interest are also
illustrated.}%
\label{pbtrans}%
\end{center}
\end{figure}

\section{Superconducting fluctuation corrections to conductivity}

We will carry out a microscopic calculation within the standard framework of
temperature diagrammatic technique for a disordered electron
system\cite{AbrikosovGD,AltshulerA1985} in the diffusive limit ($\tau^{-1}\gg
T,\alpha$). This technique has been extensively used in studying the weak
localization\cite{GorkovLK1979} and electron-electron
interaction\cite{AltshulerAL1980} corrections to the conductivity in
low-dimensional systems. In the same way as these corrections were studied
also using alternative formalisms including the non-linear sigma
model\cite{Efetov}, it seems plausible to have an alternative derivation of
the superconducting fluctuations corrections to the conductivity, which falls
in the same league. Here we will restrict ourselves solely to diagrammatic
perturbation theory.

\subsection{Basic ingredients}

Although the framework we use is standard, we will briefly discuss all the
ingredients we need mainly for two reasons. First, we want to precisely
demonstrate the way in which the presence of a pair-breaking parameter
modifies these ingredients. Second, we want to catalogue all the expressions
we need, including their limiting forms in the vicinity of the quantum phase transition.

\subsubsection{Green function}

As is standard, we assume a random disorder potential $V(\mathbf{r})$ drawn
from a Gaussian white noise ($\delta$-correlated) distribution such that
$\left\langle V(\mathbf{r})\right\rangle =0$ and $\left\langle V(\mathbf{r}%
)V(\mathbf{r}^{\prime})\right\rangle =\left\langle V^{2}\right\rangle
\delta(\mathbf{r}-\mathbf{r}^{\prime})$. In the diagrams, a dashed line
denotes
\begin{equation}
\left\langle V^{2}\right\rangle =\frac{1}{2\pi\nu\tau}%
\end{equation}
where $\nu$ is the density of states at the Fermi level and $\tau^{-1}$, as
defined earlier, is the frequency of elastic collisions. The single electron
Green function --denoted by a full line in the diagrams of Fig.
\ref{fluc_diagrams}-- is given by%
\begin{equation}
G(\omega_{n},\mathbf{p)}=\frac{1}{i\left(  \omega_{n}+\frac
{\operatorname{sign}(\omega_{n})}{2\tau}\right)  -\xi_{\mathbf{p}\ }}%
\end{equation}
where $\xi_{\mathbf{p}\ }$is the single particle excitation spectrum measured
from the chemical potential.

\subsubsection{Cooperon}

\textit{Diffuson} and \textit{Cooperon} are the key correlators, represented
by a sum over ladder diagrams involving coherent scattering by impurities, in
the particle-hole and particle-particle channel, respectively; the latter
being of interest to us in the context of superconducting fluctuations. The
expression for Cooperon --represented by a shaded rectangular block in the
diagrams-- in the presence of a pair-breaking parameter $\alpha$ is given by
\begin{equation}
C(\omega_{n},\omega_{m},\mathbf{q})={\frac{{2\pi\nu}\theta(-\omega_{n}%
\omega_{m})}{\left\vert \omega_{n}-\omega_{m}\right\vert +2\alpha_{q}}%
}\label{Cooperon1}%
\end{equation}
where $\theta(x)$ is the Heaviside theta function and%
\begin{equation}
\alpha_{q}\equiv\alpha+Dq^{2}/2
\end{equation}
Here ${\omega}_{n}$ is a fermionic Matsubara frequency and $\mathbf{q}$ is the
momentum in the effective dimension as far as the superconducting fluctuations
are concerned (note that $D$ is the diffusion constant in three dimensions as
long as the diffusion is still three-dimensional).

Coherent scattering on the same impurity, by both the electrons forming a
fluctuating Cooper pair leads to renormalization of the vertex part in the
particle-particle channel, given by%
\begin{equation}
\lambda(\omega_{n},\omega_{m},\mathbf{q})=\frac{C(\omega_{n},\omega
_{m},\mathbf{q})\ }{2\pi\nu\tau}\label{lambda}%
\end{equation}
In the diagrams shown in Fig. \ref{fluc_diagrams}, we denote $\lambda$ by a
shaded triangle.

\subsubsection{Fluctuation propagator}

The main building block of the diagrammatic technique that encodes the BCS
superconducting interaction is the so-called \textit{fluctuation propagator
}(represented by a wavy line in the diagrams)\textit{.} It is the
impurity-averaged sum over the ladder diagrams corresponding to the
electron-electron interaction in the Cooper channel. The expression in the
presence of $\alpha$, obtained using Eq. (\ref{Cooperon1}) in a standard way,
is given by
\begin{equation}
K^{-1}(\left\vert \Omega_{\nu}\right\vert ,\mathbf{q})=\ln\left(  {\frac
{T}{{T_{c0}}}}\right)  -\psi\left(  {\frac{1}{2}}\right)  +\psi\left(
{\frac{1}{2}}+{\frac{{\alpha_{q}+|\Omega}_{\nu}{|/2}}{{2\pi T}}}\right)
,\label{Pair_Correlator}%
\end{equation}
where $\Omega_{\nu}$ is a bosonic Matsubara frequency.

The pole of Eq. (\ref{Pair_Correlator}) for $q,\Omega_{v}=0$ traces the
boundary between the superconducting and normal phases (see Sec. II C). In the
limit of zero pair-breaking strength ($\alpha=0$) and near the classical
transition, $T\sim T_{c0}$, one can show that the expression for the
fluctuation propagator reduces to
\begin{equation}
K^{-1}(\left\vert \Omega_{v}\right\vert ,q)=\delta_{T}(0,T)+\frac{{Dq}%
^{2}{+|\Omega}_{\nu}{|}}{{4\pi T}}\psi^{\prime}\left(  {\frac{1}{2}}\right)
\label{KforhighT}%
\end{equation}
where $\psi^{\prime}\left(  1/2\right)  =\pi^{2}/2$ and $\delta_{T}(\alpha,T)$
is defined earlier by Eq. (\ref{deltaT})

On the other hand, at low temperatures, $T\ll\alpha_{c0},$ the fluctuation
propagator can be reduced to a form
\begin{equation}
K^{-1}(\left\vert \Omega_{\nu}\right\vert ,\mathbf{q})=\ln\left[  \frac
{\alpha_{q}+\left\vert \Omega_{\nu}\right\vert /2}{\alpha_{c}(T)}\right]
\label{KlowT}%
\end{equation}
which correctly reproduces the transition curve $\alpha=\alpha_{c}(T)$. In
certain regimes (see below) it is legitimate to expand the logarithm and get
an even simpler expression%
\begin{equation}
K(\left\vert \Omega_{\nu}\right\vert ,\mathbf{q})=\frac{1}{\delta_{\alpha
}(\alpha,T)+\frac{Dq^{2}+\left\vert \Omega_{\nu}\right\vert }{2\alpha
_{c}(T)\ }}\label{KforlowT}%
\end{equation}
where $\delta_{\alpha}(\alpha,T)$ is given by Eq. (\ref{deltaalpha}).

\subsection{Evaluation of the diagrams}

The boundary between the superconducting and normal region in the $\alpha$-$T$
phase diagram is given by Eq. (\ref{transline}). Even while superconductivity
is destroyed, superconducting fluctuations continue to persist in the normal
state region and modify the normal state conductivity. We will evaluate the
corrections to the conductivity coming from superconducting fluctuations,
using the standard Kubo formalism for linear response\cite{AbrikosovGD}. The
electromagnetic response operator $Q(\Omega_{\mu})$ is evaluated and the
external frequency is analytically continued into the upper half-plane of the
complex frequency ($i\Omega_{\mu}\rightarrow\Omega$). The fluctuation
conductivity can then be obtained using
\begin{equation}
\delta\sigma(\Omega)=\lim_{\Omega\rightarrow0}\frac{Q(-i\Omega)}{-i\Omega
}\label{cond}%
\end{equation}
once the appropriate $Q(\Omega_{\mu})$ has been evaluated.%

\begin{figure}
[ptb]
\begin{center}
\includegraphics[
trim=0.021103in 0.021113in -0.021104in -0.021114in,
height=2.2148in,
width=3.1755in
]%
{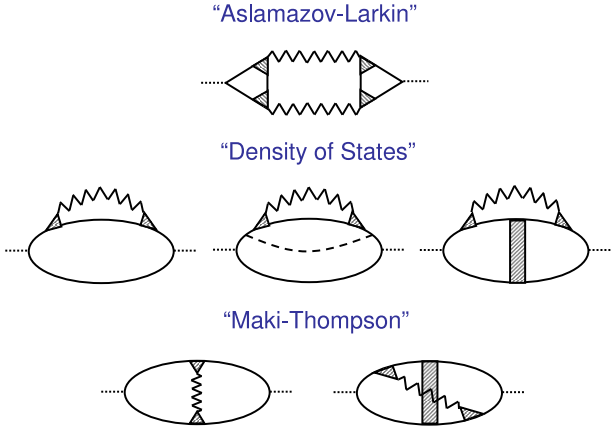}%
\caption{Diagrams for the fluctuation conductivity divided in three groups.
There is one diagram corresponding to the positive \textquotedblleft
Aslamazov-Larkin\textquotedblright\ correction. Three diagrams correspond to
the negative \textquotedblleft density-of-states\textquotedblright%
\ corrections, and each of them has two possible ways of putting arrows on the
electron Green functions. There are two diagrams corresponding to the
\textquotedblleft Maki-Thompson\textquotedblright\ interference correction
with no prescribed sign, the second of which has two ways of putting the
arrows. Full lines stand for the disorder averaged normal state Green's
function, wavy lines for the fluctuation propagator $K$, the shaded rectangles
for the Cooperon $C$ and shaded triangles for the vertex $C/2\pi\nu\tau$.}%
\label{fluc_diagrams}%
\end{center}
\end{figure}

In what follows we will evaluate $Q(\Omega_{\mu})$ using a standard set of
diagrams\cite{LarkinV}. As shown in Fig. \ref{fluc_diagrams}, these diagrams
can be divided into three groups based on their physical interpretation. Of
these, the \textquotedblleft Aslamazov-Larkin\textquotedblright\ (AL) type of
correction \cite{AslamazovL1968} is the most intuitive. It is a positive
contribution coming from additional charge transfer via fluctuating Cooper
pairs. The negative \textquotedblleft density of states\textquotedblright%
(DoS)\ correction results from the reduction of the normal single-electron
density of states at the Fermi level after accounting for the electrons
participating in fluctuating Cooper pairs. The corresponding diagrams have
only one electron line affected by the fluctuation propagator. The third, and
the more indirect, correction is given by the \textquotedblleft
Maki-Thompson\textquotedblright\ (MT)
\cite{Maki1968,Thompson1970,Thompson1971} diagrams which could be thought to
be originating from coherent Andreev scattering off the fluctuating pairs. One
can see from Fig. \ref{fluc_diagrams} that the fluctuation propagator brings
about interference between the electron lines.

Let us start by considering evaluation of the Aslamazov-Larkin diagram. The
effective triangular vertex on either side (both yield the same final
expression) is given by
\begin{align}
\Gamma(\Omega_{1\nu},\Omega_{\mu},\mathbf{q})  & =eT\sum_{\omega_{n},\text{
}\mathbf{p}}\mathbf{p}_{x}\lambda(-\omega_{n}+\Omega_{1\nu},\omega_{n}%
-\Omega_{\mu},\mathbf{q})\nonumber\\
& G(-\omega_{n}+\Omega_{1\nu},-\mathbf{p+q})G(\omega_{n}-\Omega_{\mu
},\mathbf{p)}\nonumber\\
& G(\omega_{n},\mathbf{p)}\lambda(\omega_{n},-\omega_{n}+\Omega_{1\nu
},\mathbf{q})
\end{align}
(having chosen the $x$ direction for concreteness). The presence of Heaviside
$\theta$ functions in the expression for the vertex renormalizations $\lambda
$, defined by Eqs. (\ref{Cooperon1}) and (\ref{lambda}), dictates the possible
signs and ranges of different frequencies. Taking these into account and by
making simplifications valid in the small $q$ and $\omega_{n},\Omega_{1\nu
},\Omega_{\mu}\ll\tau^{-1}$ limit of interest to us, we find a rather compact
expression
\begin{equation}
\Gamma(\Omega_{1\nu},\Omega_{\mu},\mathbf{q})=vD\mathbf{q}_{x}B(\Omega_{1\nu
},\Omega_{\mu})
\end{equation}
with%
\begin{equation}
B(\Omega_{1\nu},\Omega_{\mu})=\left[  \widetilde{\psi}(\left\vert \Omega
_{1\nu}\right\vert ,\Omega_{\mu})+\widetilde{\psi}(\left\vert \Omega_{\mu
}-\Omega_{1\nu}\right\vert ,\Omega_{\mu})\right]
\end{equation}
and
\begin{equation}
\widetilde{\psi}(w,z)\equiv\frac{1}{z}\left[  \psi\left(  {\frac{1}{2}}%
+\frac{{\alpha_{q}+}\frac{w+z}{2}}{{2\pi T}}\right)  -\psi\left(  {\frac{1}%
{2}}+\frac{{\alpha_{q}+}\frac{w}{2}}{{2\pi T}}\right)  \right]
\end{equation}
where $\psi(z)$ is the digamma function.

To evaluate the entire AL\ diagram, let us consider the summation over the
internal bosonic frequency,
\begin{equation}
I(\Omega_{\mu})\equiv T\sum_{\Omega_{1\nu}}B(\Omega_{1\nu},\Omega_{\mu}%
)^{2}K(\left\vert \Omega_{1\nu}\right\vert )K(\left\vert \Omega_{1\nu}%
-\Omega_{\mu}\right\vert )
\end{equation}
and write it as contour integration in a standard way (note that we have
temporarily suppressed the $\mathbf{q}$ dependence in the fluctuation
propagator $K\ $for the sake of compactness). By taking into account the
analyticity of the integrand, the evaluation of $I\ $is reduced to an
integration across two branch-cuts. After combining terms and analytically
continuing the external frequency to the upper half plane ($i\Omega_{\mu
}\rightarrow\Omega$), one finds that
\begin{equation}
I\equiv I_{a}+I_{b}%
\end{equation}
with%
\begin{align}
I_{a} &  =-\frac{1}{4\pi i}\frac{\Omega}{2T}\int{\frac{{d\Omega}_{1}%
}{{\mathrm{\sinh}^{2}{\frac{{\Omega}_{1}}{{2T}}}}}}K(-i\Omega_{1}%
)K(i\Omega_{1})\nonumber\\
&  \times\text{ }\left[  \widetilde{\psi}(-i\Omega_{1},-i\Omega)+\widetilde
{\psi}(i\Omega_{1},-i\Omega)\right]  ^{2}%
\end{align}
and
\begin{align}
I_{b} &  =\frac{2}{4\pi i}\int d\Omega_{1}\coth\frac{{\Omega}_{1}}{{2T}%
}K(-i\Omega_{1}-i\Omega)K(-i\Omega_{1})\nonumber\\
&  \times\text{ }\left[  \widetilde{\psi}(-i\Omega_{1}-i\Omega,-i\Omega
)+\widetilde{\psi}(-i\Omega_{1},-i\Omega)\right]  ^{2}%
\end{align}

Since the contribution to the conductivity (Eq. (\ref{cond})) goes as
$I/\Omega$, we need only consider the terms in $I\ $that are linear in
$\Omega$ since we are interested in the $\Omega\rightarrow0$ limit; terms that
are zeroth order in $\Omega$ will be cancelled by analogous terms in the
remaining diagrams to ensure the absence of anomalous diamagnetism in the
normal state. Carrying out an expansion in $\Omega$ we can write
\begin{align}
\widetilde{\psi}(-i{\Omega_{1},-}i\Omega)  & \rightarrow\frac{1}{4\pi T}%
\psi^{\prime}\left(  {\frac{1}{2}}+{\frac{{\alpha_{q}-i\Omega_{1}/2}}{{2\pi
T}}}\right) \\
& -\frac{1}{2}\frac{i\Omega}{(4\pi T)^{2}}\psi^{\prime\prime}\left(  {\frac
{1}{2}}+{\frac{{\alpha_{q}-i\Omega_{1}/2}}{{2\pi T}}}\right) \nonumber
\end{align}
Since $I_{a}$ is already linear in $\Omega$, in it we need only keep the
zeroth order term from this expansion. On the other hand, using the expansion
for both $K$ and $\widetilde{\psi}$, we have
\begin{align}
I_{b} &  =\frac{-\Omega}{2\pi}\int d\Omega_{1}\coth\frac{{\Omega}_{1}}{{2T}%
}\Bigl[4\widetilde{\psi}(-i\Omega_{1})^{2}K^{\prime}(-i\Omega_{1}%
)K(-i\Omega_{1})\nonumber\\
&  +\psi^{\prime}\left(  {\frac{1}{2}}+{\frac{{\alpha_{q}-}\frac{{i\Omega_{1}%
}}{2}}{{2\pi T}}}\right)  \psi^{\prime\prime}\left(  {\frac{1}{2}}%
+{\frac{{\alpha_{q}-}\frac{{i\Omega_{1}}}{2}}{{2\pi T}}}\right)
\frac{K(-i\Omega_{1})^{2}}{\left(  2\pi T\right)  ^{3}}\Bigr]\
\end{align}
We can integrate the first term by parts, combine similar terms from $I_{a}$
and $I_{b}$, and after some manipulation get a final expression for $I$. One
can then immediately write down the Aslamazov-Larkin fluctuation correction to
conductivity as a sum of two terms:
\begin{equation}
\delta\sigma^{AL}=\delta\sigma_{sh}^{AL}+\delta\sigma_{cth}^{AL}%
\end{equation}
where%
\begin{align}
{\delta\sigma_{sh}^{AL}}  & ={\frac{{D^{2}e^{2}}}{{2\pi Td}}}\int{\frac
{{d^{d}q}}{{(2\pi)^{d}}}\frac{{d\Omega}_{1}}{{\mathrm{\sinh}^{2}{\frac
{{\Omega}_{1}}{{2T}}}}}}\label{AL1}\\
& \Bigl[\left(  \mathrm{Im}\{K(-i\Omega_{1},q)\,\gamma(-i\Omega_{1}%
,\mathbf{q)}\}\right)  ^{2}\nonumber\\
& +\mathrm{Im}\{K(-i\Omega_{1},q)\,\gamma(-i\Omega_{1},\mathbf{q)}%
^{2}\}\mathrm{Im}\{K(-i\Omega_{1},q)\}\Bigr]\nonumber
\end{align}
and%
\begin{align}
\delta{\sigma_{cth}^{AL}} &  ={\frac{{D^{2}e^{2}i}}{{8\pi^{4}T^{3}}}}\int
\frac{{d^{d}q\,d\Omega}_{1}}{{(2\pi)^{d}}}\coth\frac{{\Omega}_{1}}{{2T}}%
q_{x}^{2}K^{2}(-i\Omega_{1},q)\nonumber\\
&  \times\psi^{\prime}\left(  {\frac{1}{2}}+{\frac{\alpha_{q}-i\Omega_{1}%
/2}{{2\pi T}}}\right)  \psi^{\prime\prime}\left(  {\frac{1}{2}}+{\frac
{\alpha_{q}{-i\Omega}_{1}{/2}}{{2\pi T}}}\right)
\end{align}
with
\begin{equation}
\gamma(-i\Omega_{1},\mathbf{q)}=\frac{\mathbf{q}}{2\pi T}\psi^{\prime}\left(
{\frac{1}{2}}+{\frac{{\alpha_{q}-i\Omega_{1}/2}}{{2\pi T}}}\right)
\label{gamma}%
\end{equation}

Going through the derivation, the reader can readily convince her/him-self
that the contribution $\delta{\sigma_{cth}^{AL}}$ would be missed if one were
to make the so-called static approximation in the effective vertex and use
$\Gamma(\Omega_{1\nu}=0,\Omega_{\mu},\mathbf{q})$ in the evaluation of the
Aslamazov-Larkin diagram. As long as one is interested in obtaining the
corrections near the classical transition, the approximation is justified and
$\delta{\sigma_{sh}^{AL}}$ is indeed the dominant contribution. We will show
below that in this limit, the result we obtain is in agreement with the
existing literature\cite{LarkinV}. In the $\alpha\ll T\sim T_{c0}$ limit, we
can expand
\begin{equation}
\gamma(-i\Omega_{1},\mathbf{q)}\rightarrow\frac{\mathbf{q}}{2\pi T}%
\psi^{\prime}\left(  {\frac{1}{2}}\right)  +\frac{{\alpha_{q}-i\Omega_{1}/2}%
}{{2\pi T}}\psi^{\prime\prime}\left(  {\frac{1}{2}}\right)
\end{equation}
where we have used ${\Omega_{1}\ll T}$, in addition. Keeping the zeroth order
term amounts to making the static approximation ($\Omega_{1}=0$) and Eq.
(\ref{AL1}) reduces to%
\begin{align}
{\delta\sigma_{sh}^{AL}}{=\frac{2}{{d}}}  & \frac{{D^{2}e^{2}}}{(2\pi T)^{3}%
}\psi^{\prime}\left(  \frac{1}{2}\right)  ^{2}\nonumber\\
& \int{\frac{{d^{d}q}}{{(2\pi)^{d}}}\frac{{d\Omega}_{1}}{{\mathrm{\sinh}%
^{2}{\frac{{\Omega}_{1}}{{2T}}}}}}q^{2}\left[  \mathrm{Im}K(-i\Omega
_{1},q)\right]  ^{2}%
\end{align}
Using Eq. (\ref{KforhighT}) for the fluctuation propagator in the limit under
consideration, we have%
\begin{equation}
\mathrm{Im}K(-i\Omega_{1},q)=\frac{{4\pi T}}{\psi^{\prime}\left(
{1/2}\right)  }\frac{\Omega_{1}}{(\widetilde{\delta}_{T}+Dq^{2})^{2}+{\Omega
}_{1}^{2}}%
\end{equation}
where
\begin{equation}
\widetilde{\delta}_{T}=\delta_{T}(0,T)\frac{{4\pi T}}{\psi^{\prime}\left(
{1/2}\right)  }%
\end{equation}
Further using $\sinh x\rightarrow x$ for small $x,$ we can write the final
expression%
\begin{align}
{\delta\sigma_{sh}^{AL}}  & {=}\ \frac{16TD^{2}{e^{2}}}{\pi d}\int
{\frac{{d^{d}qd\Omega}_{1}}{{(2\pi)^{d}}}\frac{q{^{2}}}{\left[  (\widetilde
{\delta}_{T}+Dq^{2})^{2}+{\Omega}_{1}^{2}{\mathrm{\ }}\right]  ^{2}}%
}\nonumber\\
\  & =\frac{D^{2}{e^{2}}}{T^{2}}\int{\frac{{d^{d}q}}{d{(2\pi)^{d}}}}%
\frac{{q^{2}}}{\left(  \ \widetilde{\widetilde{\delta}}_{T}+\frac{Dq^{2}}%
{2T}\right)  ^{3}}\label{ALsh_class}%
\end{align}
where we have redefined
\begin{equation}
\widetilde{\widetilde{\delta}}_{T}=\frac{4\delta_{T}(0,T)}{\pi}=\frac
{\widetilde{\delta}_{T}}{2T}%
\end{equation}
to put the integral in a form similar to what we will find in one of the
regimes near the quantum phase transition. The presentation of the results for
different dimensions will accordingly be deferred to Sec. IV in order to
facilitate the comparison.

Now we move on to the other two corrections given by the \textquotedblleft
density-of-states\textquotedblright\ and the \textquotedblleft
Maki-Thompson\textquotedblright\ diagrams as shown in Fig. \ref{fluc_diagrams}%
. The calculation follows on lines similar to what we have outlined for the
case of Aslamazov-Larkin diagram in detail above. We find that the
density-of-state fluctuation correction can be expressed as
\begin{equation}
\delta\sigma^{DoS}=\delta\sigma_{sh}^{DoS}+\delta\sigma_{cth}^{DoS}%
\end{equation}
where
\begin{align}
{\delta\sigma_{sh}^{DoS}}  & ={\frac{{D^{2}e^{2}}}{{4\pi}T{\ }}}\int
\frac{{d^{d}q\,d\omega d\Omega}}{{(2\pi)^{d}}}\tanh{\frac{{\omega}}{{2T}}%
\frac{{i}K(-i\Omega,q)}{{\mathrm{\sinh}^{2}{\frac{{\Omega}}{{2T}}}}}%
}\nonumber\\
& \left[  -\text{\textrm{Re}}{\frac{{1}}{{(\alpha_{q}-i\omega_{+})^{2}}}%
+}\text{ \textrm{Re}}{\frac{q_{x}^{2}}{{(\alpha_{q}-i\omega_{+})^{3}}}%
}\right]
\end{align}
and
\begin{equation}
{\delta\sigma_{cth}^{DoS}}=-A+\frac{3}{2}B\label{DOScth}%
\end{equation}
with%
\begin{align}
A  & =-{\frac{{De^{2}}}{{2\pi}}}\int\frac{{d^{d}q\,d\omega d\Omega}}%
{{(2\pi)^{d}}}\tanh{\frac{{\omega}}{{2T}}}\coth{{\frac{\Omega}{{2T}}}%
}\nonumber\\
& \left[  {\frac{{\ 1}}{{(\alpha_{q}-i\omega}_{+}{)^{3}}}}K(-i\Omega,q)\right]
\label{A}%
\end{align}
and
\begin{align}
B  & =-{\frac{{De^{2}}}{{2\pi}}}\int\frac{{d^{d}q\,d\omega d\Omega}}%
{{(2\pi)^{d}}}\tanh{\frac{{\omega}}{{2T}}}\coth{{\frac{\Omega}{{2T}}}%
}\nonumber\\
& \left[  \frac{q_{x}^{2}}{{(\alpha_{q}-i\omega}_{+}{)^{4}}}K(-i\Omega
,q)\right] \label{B}%
\end{align}
Here%
\begin{equation}
{\omega_{+}\equiv\omega+\Omega/2}%
\end{equation}
and $A$ and $B\ $are introduced for future convenience. The first term in each
of ${\delta\sigma_{sh}^{DoS}}$ and ${\delta\sigma_{cth}^{DoS}}$ comes from the
evaluation of the first two density-of-states diagrams, while the second term
comes from the density-of-states diagram that contains an extra Cooperon.

The \textquotedblleft Maki-Thompson\textquotedblright\ correction can be
expressed as
\begin{equation}
\delta\sigma^{MT}=\delta\sigma_{sh}^{MT}+\delta\sigma_{cth}^{MT}%
\end{equation}
where
\begin{align}
{\delta\sigma_{sh}^{MT}}  & ={\frac{{D^{2}e^{2}}}{{4\pi}T{\ }}}\int
\frac{{d^{d}q\,d\omega d\Omega}}{{(2\pi)^{d}}}\tanh{\frac{{\omega}}{{2T}}%
\frac{{i}K(-i\Omega,q)}{{\mathrm{\sinh}^{2}{\frac{{\Omega}}{{2T}}}}}%
}\nonumber\\
& \left[  {\frac{{1}}{\alpha_{q}^{2}{+\omega_{+}^{2}}}}\right]
\end{align}
while%
\begin{equation}
{\delta\sigma_{cth}^{MT}}=-A+3B\label{MTcth}%
\end{equation}
with $A$ and $B\ $defined above. The contribution ${\delta\sigma_{sh}^{MT}} $
as well as the first term in ${\delta\sigma_{cth}^{MT}}$ is obtained by
evaluating the first of the two Maki-Thompson diagrams. The evaluation of the
second diagram with an extra Cooperon yields the second term in ${\delta
\sigma_{cth}^{MT}}$ which is of the same order at low temperatures, and of
lower order at higher temperatures.

\section{Transport near the Quantum phase transition}

In the previous section we have derived the expressions for the different
fluctuation corrections to the conductivity that are valid in the vicinity of
the entire boundary (Eq. (\ref{transline})) between the superconducting and
normal phase in the $\alpha$-$T$ phase diagram. As the pair-breaking strength
is increased, superconductivity breaks down even at $T=0$ once $\alpha$
becomes equal to $\alpha_{c0}\equiv\alpha_{c}(T=0)=\pi T_{c0}/2\gamma$. In
this section we want to map out the superconducting fluctuations regimes and
obtain the finite temperature crossovers that would appear in the region
\begin{equation}%
\begin{array}
[c]{ccc}%
T & \ll & \alpha_{c}(T)\\
\alpha-\alpha_{c}(T) & \ll & \alpha_{c}(T)
\end{array}
\label{region}%
\end{equation}
near the pair-breaking quantum phase transition. Note however that the
validity of our zero temperature results will not be restricted to the
immediate neighborhood of $\alpha_{c0}$.

\subsection{Dominant fluctuation corrections}

Since in this section we are interested in the region delineated by Eq.
(\ref{region}), we will use the asymptotic form of the digamma function%
\begin{equation}
\psi\left(  {\frac{1}{2}}+{\frac{\alpha_{q}-i\Omega_{1}/2}{{2\pi T}}}\right)
\rightarrow\ln\left(  \frac{\alpha_{q}-i\Omega_{1}/2}{{2\pi T}}\right)
\label{digamma_asymp}%
\end{equation}
in the forthcoming treatment. Let us begin by analyzing the Aslamazov-Larkin
correction. Using Eq. (\ref{digamma_asymp}) inside the expression for $\gamma$
as defined by Eq. (\ref{gamma}), we obtain a reduced form%
\begin{equation}
\gamma(-i\Omega_{1},\mathbf{q})\rightarrow\frac{\mathbf{q}}{{\alpha
_{q}-i\Omega_{1}/2}}\rightarrow\frac{\mathbf{q}}{{\alpha}_{c}(T){\ }}%
\end{equation}
where the second limit is justified \textit{a posteriori} by the behavior of
the integral below. We have
\begin{equation}
{\delta\sigma_{sh}^{AL}=\frac{{D^{2}e^{2}}}{{\pi Td\alpha}_{c}^{2}}}\int
{\frac{{d^{d}q}}{{(2\pi)^{d}}}\frac{{d\Omega}_{1}}{{\mathrm{\sinh}^{2}%
{\frac{{\Omega}_{1}}{{2T}}}}}}q^{2}\left[  \mathrm{Im}K(-i\Omega
_{1},q)\right]  ^{2}\label{AL1_QPT}%
\end{equation}
where
\begin{equation}
\mathrm{Im}K(-i\Omega_{1},q)=\frac{\Omega_{1}/(2\alpha_{c}(T))}{\left(
\delta_{\alpha}(\alpha,T)+\frac{Dq^{2}}{2\alpha_{c}(T)\ }\right)  ^{2}+\left(
\frac{\Omega_{1}}{2\alpha_{c}(T)\ }\right)  ^{2}}%
\end{equation}
is obtained using Eq. (\ref{KforlowT}). Indeed the integral is dominated by
small values of $q$ and $\Omega_{1}$ and ${\delta\sigma_{sh}^{AL}}$ is
critical in the vicinity of $\alpha_{c}(T)$. That is also the reason why it
was legitimate to use the reduced form for $K$.

At this point the reader should observe that to analytically analyze
${\delta\sigma_{sh}^{AL}}$ any further, we need to order the energy scales,
$T\ $and $\alpha-\alpha_{c}(T)$. The frequency integral can then be carried
out to get the limiting form%
\begin{equation}
{\delta\sigma_{<,sh}^{AL}=}\frac{2\pi}{3}\frac{D^{2}{e^{2}}T^{2}}{\alpha
_{c}^{4}}\int{\frac{{d^{d}q}}{d{(2\pi)^{d}}}}\frac{q^{2}}{\left(
\delta_{\alpha}(\alpha,T)+\frac{Dq^{2}}{2\alpha_{c}(T)\ }\right)  ^{4}%
}\label{ALsh<}%
\end{equation}
for $T\ll\alpha-\alpha_{c}(T)\ $and%
\begin{equation}
{\delta\sigma_{>,sh}^{AL}}=\frac{D^{2}{e^{2}}T}{\alpha_{c}^{3}}\int
{\frac{{d^{d}q}}{d{(2\pi)^{d}}}}\frac{q^{2}\ }{\left(  \delta_{\alpha}%
(\alpha,T)+\frac{Dq^{2}}{2\alpha_{c}(T)\ }\right)  ^{3}}\label{ALsh>}%
\end{equation}
for $T\gg\alpha-\alpha_{c}(T)$. Observe that the integrand in Eq.
(\ref{ALsh>}) has exactly the same form as the integrand in Eq.
(\ref{ALsh_class}) corresponding to the Aslamazov-Larkin correction in the
temperature vicinity above the classical transition at $T_{c0}$ in the absence
of any pair-breaking perturbation. Whereas $\delta_{T}(0,T)$ and $T\ $were the
parameters in that integrand, here they are $\delta_{\alpha}(\alpha,T)$ and
$\alpha_{c}(T)\ $instead.

To be able to access the crossover between the two limiting forms,
${\delta\sigma_{<,sh}^{AL}}$ and ${\delta\sigma_{>,sh}^{AL}}$, we recast Eq.
(\ref{AL1_QPT}) in terms of dimensionless variables to obtain
\begin{equation}
{\delta\sigma_{sh}^{AL}=\frac{2{D^{2}e}^{2}}{{\pi dT}^{2}}}\left(  \frac
{2T}{D}\right)  ^{\frac{2+d}{2}}F(\eta)\label{ALshSF}%
\end{equation}
where
\begin{equation}
F(\eta)=\int{\frac{{d^{d}q}}{{(2\pi)^{d}}}\frac{{d\Omega}_{1}}{\mathrm{\sinh
}{^{2}\Omega}_{1}}}\frac{q^{2}\Omega_{1}^{2}}{\left[  \left(  \ \eta
+q^{2}\right)  ^{2}+\Omega_{1}{}^{2}\right]  ^{2}}\label{ScalFunc}%
\end{equation}
is a scaling function of the dimensionless parameter
\begin{equation}
\eta\equiv\frac{\alpha-\alpha_{c}(T)}{T}\label{eta}%
\end{equation}
In the $\eta\gg1$ limit, ${\delta\sigma_{sh}^{AL}}$ reduces to Eq.
(\ref{ALsh<}) while in the $\eta\ll1$ limit it reduces to Eq. (\ref{ALsh>}).
The numerical evaluation of $F(\eta)$ gives the behavior ${\delta\sigma
_{sh}^{AL}}$ over the entire range of $\eta$.

Now let us proceed to analyze the ${\delta\sigma_{cth}^{AL}}$ part of the
Aslamazov-Larkin correction. After using the asymptotic form of the digamma
function given by Eq. (\ref{digamma_asymp}) to get a simplified expression, we
express ${\delta\sigma_{cth}^{AL}}$ as a sum of two terms:
\begin{align}
\delta{\sigma_{cth}^{AL}} &  ={\frac{{D^{2}e^{2}}}{{\pi i}}}\int\frac
{{d^{d}q\,d\Omega}_{1}}{{(2\pi)^{d}}}\frac{q_{x}^{2}K^{2}(-i\Omega_{1}%
,q)}{(\alpha_{q}-i\Omega_{1}/2)^{3}}\nonumber\\
&  \times\left[  \left(  \coth\frac{{\Omega}_{1}}{{2T}}-\operatorname{sign}%
\frac{{\Omega}_{1}}{{2T}}\right)  +\operatorname{sign}\frac{{\Omega}_{1}}%
{{2T}}\right] \label{AL2lowT}%
\end{align}
At any finite temperature the first (difference)\ term can be shown to be
sub-dominant as compared to $\delta{\sigma_{sh}^{AL}}$ while for $T=0$ it
vanishes identically. Thus we need evaluate only the second term which is
independent of temperature. To this end, we carry out a bunch of manipulations
and do the frequency integration by parts, to obtain a neat expression which
we denote by
\begin{equation}
{\delta{\sigma_{0,cth}^{AL}}=\frac{4{De^{2}}}{{\pi d(d+2)}}}\int\frac
{{d^{d}q\ }}{{(2\pi)^{d}}}\frac{(Dq^{2})^{2}K^{2}(0,q)}{\alpha_{q}^{3}%
}\label{AL2const}%
\end{equation}

As for the Maki-Thompson and the density-of-states corrections, we find that
$\delta\sigma_{sh}^{DoS}+\delta\sigma_{sh}^{MT}$ is sub-dominant as compared
to the Aslamazov-Larkin correction ${\delta{\sigma_{sh}^{AL}}}$. On the other
hand, the same way as we did in Eq. (\ref{AL2lowT}), we can express
$\delta\sigma_{cth}^{DoS}$ and $\delta\sigma_{cth}^{MT}$ as a sum of a
temperature-dependent term that vanishes at $T=0$ and a second
temperature-independent term. The temperature-dependent term is again
sub-dominant as compared to ${\delta{\sigma_{sh}^{AL}}}$ and after some
manipulation, we find the temperature-independent part of $A$ and $B$ (see
Eqs. (\ref{A}) and (\ref{B})) to be given by
\begin{equation}
A_{0}={\frac{2{De^{2}}}{{\pi}}}\int\frac{{d^{d}q\ }}{d{(2\pi)^{d}}}%
\frac{Dq^{2}K(0,q)}{\alpha_{q}^{2}}%
\end{equation}
and
\begin{equation}
B_{0}={\frac{4{De^{2}}}{3{\pi}}}\int\frac{{d^{d}q\ }}{{d(d+2)(2\pi)^{d}}}%
\frac{(Dq^{2})^{2}K(0,q)}{\alpha_{q}^{3}}%
\end{equation}
We can re-express $B_{0}\ $as
\begin{align}
B_{0}  & =\frac{A_{0}}{3}-\widetilde{B}_{0}\\
\widetilde{B}_{0}  & ={\frac{2{De^{2}}}{3{\pi d(d+2)}}}\int\frac{{d^{d}q\ }%
}{{(2\pi)^{d}}}\frac{(Dq^{2})^{2}K^{2}(0,q)}{\alpha_{q}^{3}}%
\end{align}
such that we have
\begin{align}
{\delta{\sigma_{0,cth}^{AL}}}  & {=}{6}\widetilde{B}_{0}\\
{\delta{\sigma_{0,cth}^{DoS}}}  & {=}{-}\frac{{A}_{0}+3\widetilde{B}_{0}}{2}\\
{\delta{\sigma_{0,cth}^{MT}}}  & {=}{-3}\widetilde{B}_{0}%
\end{align}
to yield
\begin{align}
{\delta{\sigma_{0,cth}}}  & {\equiv}{}\text{ }{\delta{\sigma_{0,cth}^{AL}%
+}\delta{\sigma_{0,cth}^{DoS}+}\delta{\sigma_{0,cth}^{MT}}}\nonumber\\
& =-\frac{A_{0}-3\widetilde{B}_{0}}{2}=-\frac{3B_{0}}{2}\label{sigmazeroT}\\
& ={\frac{-2{De^{2}}}{{\pi}}}\int\frac{{d^{d}q\ }}{{d(d+2)(2\pi)^{d}}}%
\frac{(Dq^{2})^{2}K(0,q)}{\alpha_{q}^{3}}\nonumber
\end{align}
where $A_{0},$ $B_{0}$ and $\widetilde{B}_{0}$ are all positive quantities
defined above. The expression for $K(0,q)$ is given by Eq. (\ref{KlowT}) which
is appropriate for $T\ll\alpha_{c0}$.$\ $

\subsection{Fluctuation regimes in the $\alpha-T$ plane}%

\begin{figure}
[ptb]
\begin{center}
\includegraphics[
height=2.4358in,
width=3.4123in
]%
{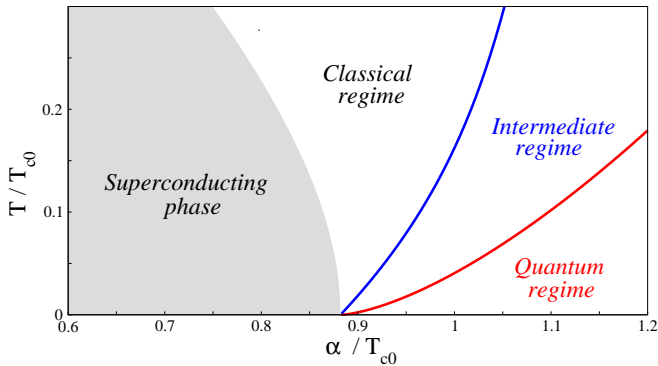}%
\caption{Phase diagram showing the vicinity of the superconducting quantum
critical point realized via a pair-breaking quantum phase transition as
illustrated in Fig. \ref{pbtrans}. The boundary between the classical and the
intermediate regime is given by $T=\alpha-\alpha_{c}(T).$ The analytical
estimate for the boundary between quantum and intermediate regimes is given by
Eqs. (\ref{1dQIbound}), (\ref{2dQIbound}), and (\ref{3dQIbound}) for the case
of a nanowire or doubly-connected cylinder, a thin film and a bulk
superconductor, respectively. For clarity, only the case of $d=1$ is shown.
The quantum regime extends to higher temperatures as the effective
dimensionality of the system is increased. }%
\label{phase_diagram}%
\end{center}
\end{figure}

In the previous subsection we have identified and analyzed the dominant
fluctuation corrections to the normal state conductivity in the vicinity of a
pair-breaking quantum phase transition out of a superconducting state. Based
on this analysis, the fluctuation conductivity is%
\begin{equation}
\delta\sigma(\alpha,T)=\delta\sigma_{0,cth}(\alpha)+\delta\sigma_{sh}%
^{AL}(\alpha,T)\label{sigmatot}%
\end{equation}
where $\delta\sigma_{0,cth}(\alpha)$ is given by Eq. (\ref{sigmazeroT}) and
$\delta\sigma_{sh}^{AL}(\alpha,T)$ is given by Eq. (\ref{AL1_QPT}). At the
absolute zero of the temperature, the correction to the conductivity is given
by%
\begin{equation}
\delta\sigma(\alpha,0)=\delta\sigma_{0,cth}(\alpha)
\end{equation}
and it continues to be the dominant correction up to a temperature scale
$T_{0}(\alpha)$ at which $\delta\sigma_{sh}^{AL}(\alpha,T)$ becomes
comparable. This regime with a negative fluctuation correction which turns out
to be almost non-critical, is the \textquotedblleft quantum
regime\textquotedblright\ with thermal fluctuations playing no role
whatsoever. In this regime, the Maki-Thompson correction turns out to be
negative and is half the magnitude of the positive Aslamazov-Larkin
correction; the negative density-of-states correction too is of the same
order, and the total is negative. The presence of the superconducting
fluctuations thus impedes the flow of current contrary to the naive
expectation. It is important to note that the quantum regime is outside the
scope of any theoretical approach that does not take into account the electron
degrees of freedom in addition to the superconducting fluctuations.

On the other hand, the regime defined by $T>\alpha-\alpha_{c}(T)$ is dominated
by the Aslamazov-Larkin correction ${\delta\sigma_{sh}^{AL}}$ in its limiting
form ${\delta\sigma_{>,sh}^{AL}}$ given by Eq. (\ref{ALsh>}); this is the
\textquotedblleft classical regime\textquotedblright. The \textquotedblleft
intermediate regime\textquotedblright\ in between the classical and quantum
regimes is dominated by ${\delta\sigma_{<,sh}^{AL}}$, which is the limiting
form of the Aslamazov-Larkin correction appropriate for $T\ll\alpha-\alpha
_{c}(T)$, and is given by Eq. (\ref{ALsh<}). The fluctuation conductivity in
both the classical and the intermediate regime enhances the normal state
conductivity due to the additional channel of transport via the fluctuating
Cooper pairs and is more critical than that in the quantum regime.

The presence of these three regimes means that the conductivity behavior
depends on how the quantum phase transition or the low temperature transition
is approached. If the quantum critical point is approached by coming down in
temperature at fixed value of $\alpha=\alpha_{c0}$, then the measurement
trajectory sweeps exclusively across the classical regime and diverges as the
temperature tends to zero. On the other hand if $\alpha_{c0}$ is approached by
tuning the pair-breaking parameter at a fixed value of $T=0$ then the path
lies entirely in the quantum regime showing the characteristic increase in
resistance. On the contrary, going away from $\alpha_{c0}$ results in a
decrease in the resistance as $\alpha$ is increased; if the pair-breaking
strength were to be tuned by a parallel magnetic field, a negative
magnetoresistance is observed. The intermediate regime is swept only while the
measurement trajectory crosses over from the quantum to the classical regime
or vice-versa. A trajectory in which the temperature is varied at a fixed
value of $\alpha>\alpha_{c0}$ or $\alpha$ is tuned at a non-zero temperature,
results in a non-trivial conductivity with a non-monotonic behavior.

\subsubsection{Nanowire or doubly-connected cylinder (d = 1)}

Consider a pair-breaking quantum phase transition in a superconducting
nanowire driven by tuning a magnetic field, the concentration of magnetic
impurities, or yet another pair-breaking perturbation. As long as the diameter
of the wire is smaller than the superconducting coherence length, the system
is effectively one-dimensional as far as superconducting fluctuations are
concerned. One could also consider a doubly-connected cylinder instead of a
wire, as discussed in Sec. II.

For $d=1$ the fluctuation conductivity in the classical regime ($T>\alpha
-\alpha_{c}(T)$), given by Eq. (\ref{ALsh>}), can be written in terms of a
dimensionless integral to obtain
\begin{align}
{\delta\sigma_{>,sh}^{AL}}(\alpha,T)  & =\frac{\sqrt{D}{e^{2}}T\ }{2\pi
\alpha_{c}^{3/2}}\int_{-\infty}^{\infty}dx\frac{x^{2}}{\left(  \delta_{\alpha
}(\alpha,T)+x^{2}/2\right)  ^{3}}\nonumber\\
& =\frac{\sqrt{D}{e^{2}}}{4\sqrt{2}}\frac{T}{\ (\alpha-\alpha_{c}(T))^{3/2}%
}\label{classregd1}%
\end{align}
The temperature dependence is more revealing if one uses $\alpha_{c}(T)\sim$
$\alpha_{c0}-\pi^{2}T^{2}/(6\alpha_{c0})$ in the vicinity of $\alpha_{c0}$
(see Eq. (\ref{alphaexp})). In particular, if the quantum critical point at
$\alpha_{c0}$ is approached by coming down in temperature, then
\begin{equation}
{\delta\sigma_{>,sh}^{AL}}(\alpha_{c0},T)=\frac{3\sqrt{3}\sqrt{D}{e^{2}}}%
{2\pi^{3}}\frac{\alpha_{c0}^{3/2}}{\ T^{2}\ }\label{QCdivd1}%
\end{equation}
shows a quantum critical divergence with the power $T^{-2}$ when
$T\rightarrow0$.

It is instructive to compare the result in Eq. (\ref{classregd1}) in the
classical regime with the fluctuation conductivity
\begin{equation}
{\delta\sigma_{sh}^{AL}}(0,T)=\frac{\pi\sqrt{\pi}\sqrt{D}e^{2}}{32\sqrt{2}%
}\frac{T_{c0}}{(T-T_{c0})^{3/2}\ }%
\end{equation}
in the temperature vicinity above the classical transition at $T_{c0}$ in the
absence of any pair-breaking perturbation. Derivation of this result requires
evaluating Eq. (\ref{ALsh_class}) which contains the same dimensionless
integral as in the classical regime near the quantum phase transition.

By using Eq. (\ref{ALsh<}) we have
\begin{align}
{\delta\sigma_{<,sh}^{AL}(\alpha,T)}  & =\frac{\sqrt{D}T^{2}e^{2}}{3\alpha
_{c}^{5/2}}\ \int_{-\infty}^{\infty}dx\frac{x^{2}}{\left(  \delta_{\alpha
}(\alpha,T)+x^{2}/2\right)  ^{4}}\nonumber\\
& =\frac{\pi\sqrt{D}{e^{2}}}{12\sqrt{2}}\frac{T^{2}}{\ (\alpha-\alpha
_{c}(T))^{5/2}}%
\end{align}
in the intermediate regime between the classical and quantum regimes. The
fluctuation conductivity contains an additional power of $T/(\alpha-\alpha
_{c}(T))$ as compared to the classical regime.

Expressed in terms on the scaling function that we introduced in Eq.
(\ref{ALshSF}) we have
\begin{equation}
{\delta\sigma_{sh}^{AL}(\alpha,T)=\frac{4\sqrt{2}{\sqrt{D}e}^{2}}{{\pi}%
\sqrt{T}}}F(\eta)
\end{equation}
with
\begin{equation}
F(\eta)=\left\{
\begin{array}
[c]{cccc}%
\frac{\pi}{32}\eta^{-3/2} &  &  & \eta\ll1\\
\ \frac{\pi^{2}}{96}\ \eta^{-5/2} &  &  & \eta\gg1
\end{array}
\right.
\end{equation}
where $\eta$ is defined in Eq. (\ref{eta}). The full scaling function given by
Eq. (\ref{ScalFunc}) and the negative temperature independent term
$\delta\sigma_{0,cth}(\alpha)$, as discussed below, can be evaluated
numerically. In this way, we are able to get the behavior of the fluctuation
conductivity as a function of temperature and pair-breaking parameter, in the
entire neighborhood of the superconducting quantum critical point. The exact
boundary of the quantum regime in the $\alpha$-$T$ phase diagram can then be
identified by tracing the curve on which the fluctuation conductivity becomes
zero while changing sign from positive to negative.%

\begin{figure}
[ptb]
\begin{center}
\includegraphics[
height=4.4096in,
width=3.1273in
]%
{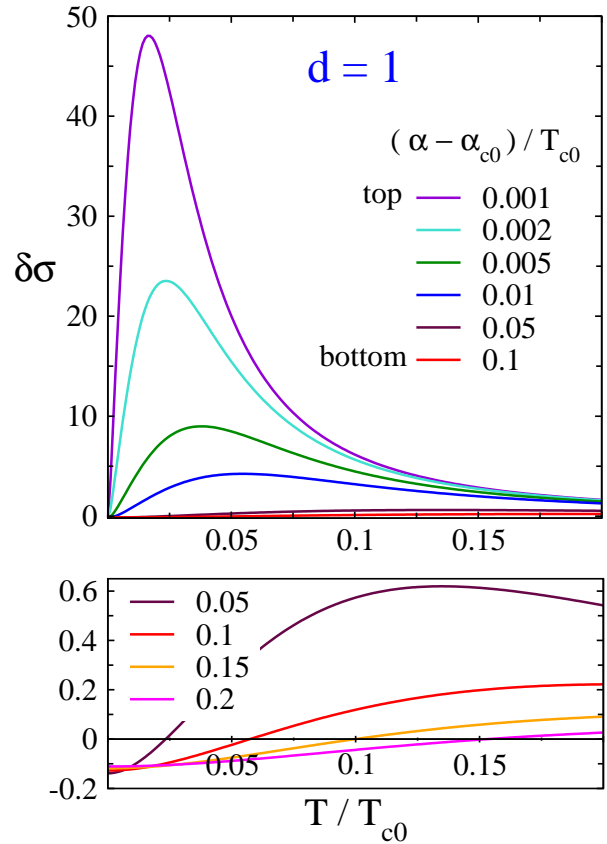}%
\caption{Temperature dependence of the fluctuation correction to the
conductivity (in units of $\sqrt{D}e^{2}$) in the vicinity of a pair-breaking
quantum phase transition at $\alpha=\alpha_{c0}=0.889T_{c0}$ for $d=1$
(nanowire or doubly-connected cylinder). Plots are shown for different values
of $(\alpha-\alpha_{c0})/T_{c0}.$ When the superconducting QCP is approached
by lowering the temperature at $\alpha=\alpha_{c0}$, the quantum critical
divergence of the conductivity is given by Eq. (\ref{QCdivd1}) (not included
in the figure). The lower panel clearly displays the temperatures
$T_{0}(\alpha)\ $(analytically estimated by Eq. (\ref{1dQIbound}), at which
the correction becomes negative, there by signalling the entry into the
quantum regime at a given value of $\alpha$.}%
\label{1dTplot}%
\end{center}
\end{figure}

The temperature independent fluctuation correction which dominates the quantum
regime is given by evaluating Eq. (\ref{sigmazeroT}), to obtain
\begin{equation}
\delta\sigma_{0,cth}(\alpha)=-\frac{\sqrt{D}{e^{2}}}{3\pi^{2}\sqrt{\alpha
_{c0}}}\int_{-\infty}^{\infty}{dx}\frac{x^{4}}{h_{\alpha}(x)^{3}\ln h_{\alpha
}(x)}%
\end{equation}
where
\begin{equation}
h_{\alpha}(x)\equiv1+\delta_{\alpha}(\alpha,0)+x^{2}/2\label{h}%
\end{equation}
The correction has no critical divergence at $\alpha=\alpha_{c0}$ (i.e.
$\delta_{\alpha}(\alpha,0)=0$) and even the first derivative is non-critical.
The second derivative diverges as $(\alpha-\alpha_{c0})^{-1/2}$ thereby giving
the first non-analytic term in the expansion (obtained by integrating twice)
around $\alpha_{c0}$. We thus obtain%
\begin{align}
& \delta\sigma_{0,cth}(\alpha)-\delta\sigma_{0,cth}(\alpha_{c0})\\
& =\frac{e^{2}\sqrt{D}}{\sqrt{\alpha_{c0}}}\left[  a_{1}\text{ }\delta
_{\alpha}(\alpha,0)\ +b_{1}\delta_{\alpha}(\alpha,0)^{3/2}+...\right]
\nonumber
\end{align}
with $a_{1}=0.386$ and $b_{1}=-4\sqrt{2}/(3\pi)$. By using the first term in
the expansion, one can analytically estimate the boundary between the
intermediate and the quantum regimes to be
\begin{equation}
\frac{T_{0}(\alpha)}{T_{c0}}\sim\left(  \frac{\alpha-\alpha_{c0}}{\alpha_{c0}%
}\right)  ^{7/4}\label{1dQIbound}%
\end{equation}

\begin{figure}
[ptb]
\begin{center}
\includegraphics[
height=4.2583in,
width=3.1007in
]%
{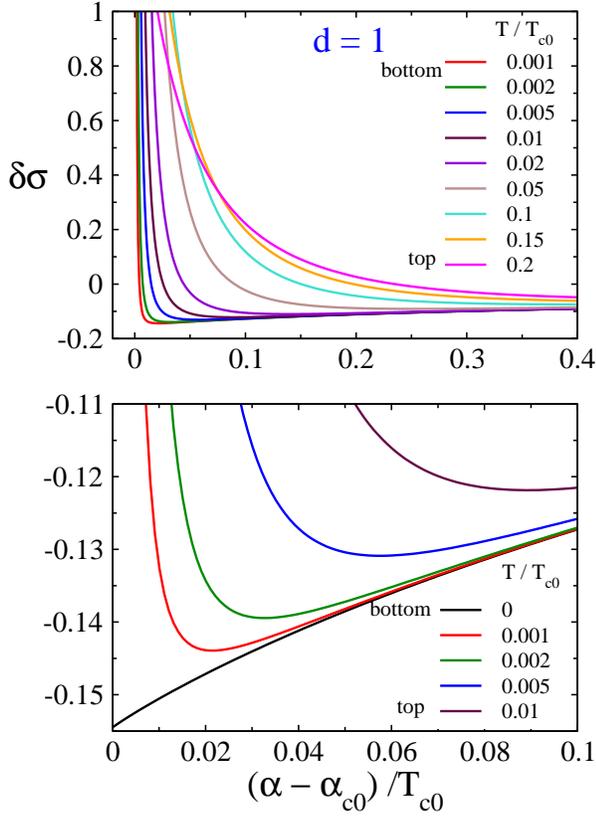}%
\caption{Fluctuation correction to the conductivity (in units of $\sqrt
{D}e^{2}$) in the vicinity of a pair-breaking quantum phase transition at
$\alpha_{c0}=0.889T_{c0}$ for $d=1$ (nanowire or doubly-connected cylinder).
Plots show the behavior of $\delta\sigma$ as one sweeps $(\alpha-\alpha
_{c0})/T_{c0}$ at a given value of $T/T_{c0}.$The lower panel shows the
negative correction at $T=0$ and a clear upturn of the conductivity which is
characteristic of the quantum regime. This would correspond to a decrease in
resistance (a negative magnetoresistance, if $\alpha$ is tuned by a magnetic
field) as $\alpha$ is increased at $T=0$ and a clearly visible non-monotonic
behavior at low temperatures.}%
\label{1dalphaplot}%
\end{center}
\end{figure}

In Figs. \ref{1dTplot} and \ref{1dalphaplot} we display the plots for the
fluctuation conductivity when the vicinity of the pair-breaking quantum phase
transition is explored either by sweeping the temperature or the pair-breaking
parameter (see the figure captions for details). As discussed above, a quantum
critical divergence is expected as temperature is lowered by sitting at
$\alpha=\alpha_{c0}$ and is not shown in the figure. If $\alpha$ is increased
starting from $\alpha_{c0}$ at a fixed value of $T=0$, the conductivity
increases monotonically, correspondingly giving a decrease in resistance i.e.
a negative magnetoresistance if $\alpha$ is tuned by a magnetic field. The
conductivity shows a non-monotonic behavior as $\alpha$ is varied at a fixed
$T\neq0$ or $T\ $is varied at a fixed $\alpha>\alpha_{c0}$, the behavior being
more distinct for smaller values of $T$ and $\alpha-\alpha_{c0}$, respectively.%

\subsubsection{Thin film (d = 2)}

Consider a pair-breaking quantum phase transition in a superconducting thin
film whose thickness is smaller than the coherence length which makes the
system effectively two-dimensional as far as superconducting fluctuations are
concerned. The transition can be driven e.g. by tuning a pair-breaking
perturbation such as a parallel magnetic field or the concentration of
magnetic impurities as discussed in Sec. II.

The remarkable fact about the fluctuation conductivity in two dimensions is
that it is an universal quantity, independent of the properties of the
material under consideration. The fluctuation conductivity in the classical
regime ($T>\alpha-\alpha_{c}(T)$), given by Eq. (\ref{ALsh>}), can be
evaluated for $d=2$ to obtain
\begin{align}
{\delta\sigma_{>,sh}^{AL}}(\alpha,T)  & =\frac{Te^{2}}{4\pi\alpha_{c}}\int
_{0}^{\infty}dx\frac{x^{3}}{\left(  \delta_{\alpha}(\alpha,T)+x^{2}/2\right)
^{3}}\ \nonumber\\
& =\frac{{e^{2}}}{4\pi}\frac{T}{\alpha-\alpha_{c}(T)\ }\label{classregd2}%
\end{align}
with $\alpha_{c}(T)\sim$ $\alpha_{c0}-\pi^{2}T^{2}/(6\alpha_{c0})$ in the
vicinity of $\alpha_{c0}$ (see Eq. (\ref{alphaexp})). If the quantum critical
point is approached by lowering the temperature at a fixed $\alpha=\alpha
_{c0}$, then we have
\begin{equation}
{\delta\sigma_{>,sh}^{AL}}(\alpha_{c0},T)=\frac{3{e^{2}}}{2\pi^{3}}%
\frac{\alpha_{c0}}{\ T\ }\label{QCdivd2}%
\end{equation}
which shows a quantum critical divergence $T^{-1}$ when $T\rightarrow0$.

The fluctuation conductivity Eq. (\ref{ALsh_class}) in the temperature
vicinity above the classical transition at $T_{c0}$ in the absence of any
pair-breaking perturbation, contains the same form of the integrand as in the
classical regime near the quantum phase transition therefore giving a result
\begin{equation}
{\delta\sigma_{sh}^{AL}}(0,T)=\frac{e^{2}}{16}\frac{T_{c0}}{T-T_{c0}}%
\end{equation}
analogous to Eq. (\ref{classregd2}) and consistent with the classic
literature\cite{LarkinV}.

By using Eq. (\ref{ALsh<}) the fluctuation conductivity in the intermediate
regime is given by
\begin{align}
{\delta\sigma_{<,sh}^{AL}(\alpha,T)}  & {=}\frac{T^{2}e^{2}}{12\alpha_{c}^{2}%
}\int_{0}^{\infty}dx\frac{x^{3}}{\left(  \delta_{\alpha}(\alpha,T)+x/2\right)
^{4}}\nonumber\\
& =\frac{{e^{2}}}{18}\frac{T^{2}}{(\alpha-\alpha_{c}(T))^{2}}%
\end{align}
It contains an additional power of $T/(\alpha-\alpha_{c}(T))$ as compared to
the classical regime, just as we found for $d=1$. Expressed in terms on the
scaling function that we introduced in Eq. (\ref{ALshSF}) we have
\begin{equation}
{\delta\sigma_{sh}^{AL}(\alpha,T)=\ \frac{4{e}^{2}}{{\pi}}}\ F(\eta)
\end{equation}
with
\begin{equation}
F(\eta)=\left\{
\begin{array}
[c]{cccc}%
\frac{1}{16}\eta^{-1} &  &  & \eta\ll1\\
\ \frac{\pi}{72}\ \eta^{-2} &  &  & \eta\gg1
\end{array}
\right.
\end{equation}
and the full form of $F(\eta)$ can be evaluated by doing the integrals in Eq.
(\ref{ScalFunc}) numerically.%

\begin{figure}
[ptb]
\begin{center}
\includegraphics[
height=4.2833in,
width=3.0832in
]%
{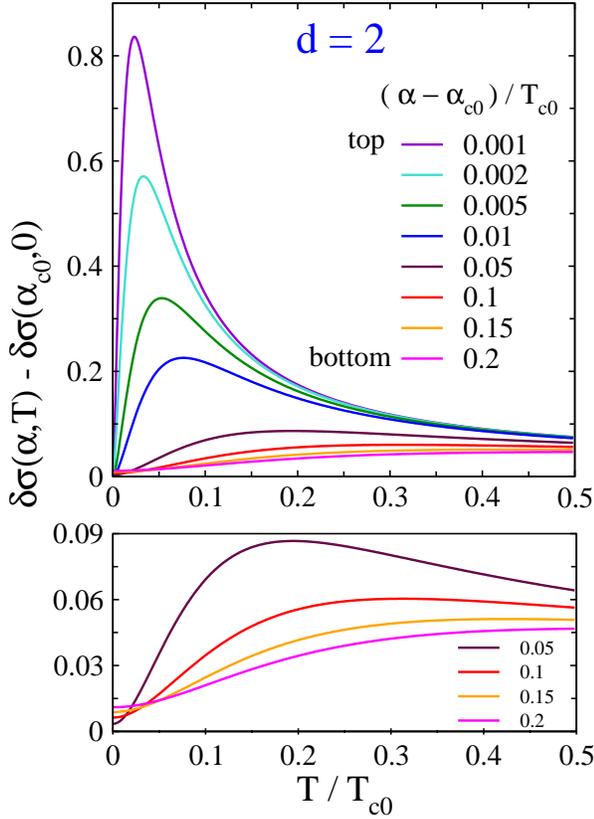}%
\caption{Temperature dependence of the fluctuation correction to the
conductivity (in units of $e^{2}$) in the vicinity of a pair-breaking quantum
phase transition at $\alpha=\alpha_{c0}=0.889T_{c0}$, for the case of a thin
film ($d=2$). Plots are shown for different values of $(\alpha-\alpha
_{c0})/T_{c0}$. When the superconducting QCP is approached by lowering the
temperature at $\alpha=\alpha_{c0}$, the quantum critical divergence of the
conductivity is given by Eq. (\ref{QCdivd2}) (not included in the figure).
Note that we have plotted the difference $\delta\sigma(\alpha,T)-\delta
\sigma(\alpha_{c0},0)$ which is always positive given that $\delta
\sigma(\alpha,0)$ is most negative at $\alpha=\alpha_{c0}$ (see Fig.
\ref{2dalphaplot}).}%
\label{2dTplot}%
\end{center}
\end{figure}

The temperature independent quantum correction which dictates the behavior in
the quantum regime can be obtained by using Eq. (\ref{sigmazeroT}) to have
\begin{equation}
\delta\sigma_{0,cth}(\alpha)=-{\frac{{e^{2}}}{8{\pi}^{2}}}\int_{0}^{\infty
}{dx}\frac{x^{5}}{h_{\alpha}(x)^{3}\ln h_{\alpha}(x)}%
\end{equation}
where $h_{\alpha}(x)$ is defined by Eq. (\ref{h}). For large $x$ the integrand
goes as $1/(x\ln x)$ and the integral has a very weak ultraviolet divergence
which can be isolated by evaluating the integral analytically by parts
($\Lambda$ is the cut-off). Indeed at $\alpha=\alpha_{c0}$ we find
\begin{align}
\delta\sigma_{0,cth}(\alpha_{c0})  & =-\frac{e^{2}}{2\pi^{2}}\ln\left(
\ln\frac{\Lambda}{2}\right) \nonumber\\
& +\frac{e^{2}}{4\pi^{2}}\int_{0}^{\infty}{dx\frac{{x\ \ln(\ln(1+x/2)\ }%
}{{\,\left(  1+x/2\right)  ^{3}\ }}}\nonumber\\
& =-\frac{e^{2}}{2\pi^{2}}\ln\left(  \ln\frac{\Lambda}{2b}\right)
\end{align}
where $b=1.12292$.$\ $We can henceforth thus consider the difference
$\delta\sigma_{0,cth}(\alpha)-\delta\sigma_{0,cth}(\alpha_{c0})$ that has a
convergent integral to be evaluated numerically.

Here again as in the $d=1$ case, $\delta\sigma_{0,cth}(\alpha)$ as well as the
first derivative has no critical divergence at $\alpha=\alpha_{c0}$. From the
divergence in the second derivative, we can get the first non-analytic term in
the expansion around $\alpha_{c0}$. We thus have
\begin{align}
& \delta\sigma_{0,cth}(\alpha)-\delta\sigma_{0,cth}(\alpha_{c0})\\
& =e^{2}\left[  a_{2}\text{ }\delta_{\alpha}(\alpha,0)\ +b_{2}\delta_{\alpha
}(\alpha,0)^{2}\ln\delta_{\alpha}(\alpha,0)\ +...\right] \nonumber
\end{align}
with $a_{2}=0.070$ and $b_{2}=1/(2\pi^{2})$. By using the first term in the
expansion, one can analytically estimate the boundary between the intermediate
and the quantum regimes to be
\begin{equation}
\frac{T_{0}(\alpha)}{T_{c0}}\sim\left(  \frac{\alpha-\alpha_{c0}}{\alpha_{c0}%
}\right)  ^{3/2}\label{2dQIbound}%
\end{equation}

\begin{figure}
[ptb]
\begin{center}
\includegraphics[
height=4.147in,
width=3.0383in
]%
{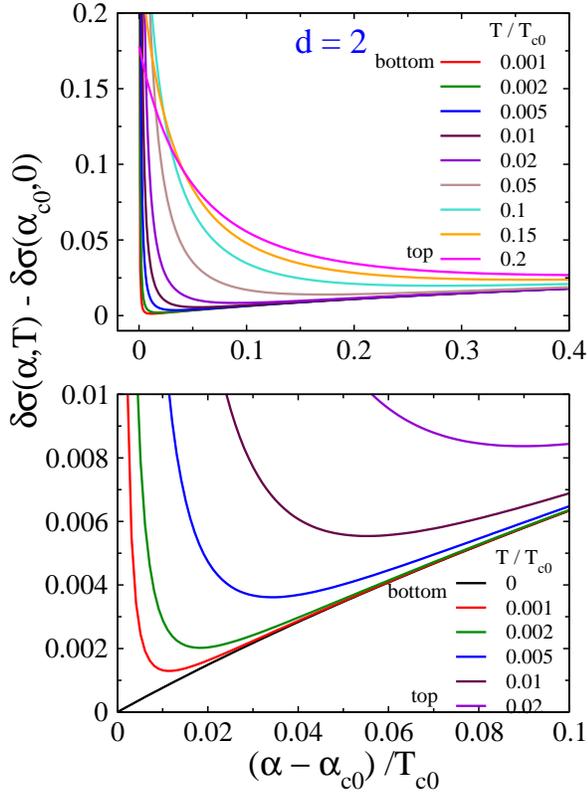}%
\caption{Fluctuation correction to the conductivity (in units of $e^{2}$) in
the vicinity of a pair-breaking quantum phase transition at $\alpha
_{c0}=0.889T_{c0}$ for $d=2$ (thin films). Plots show the behavior of
$\delta\sigma(\alpha,T)-\delta\sigma(\alpha_{c0},0)$ as one sweeps
$(\alpha-\alpha_{c0})/T_{c0}$ at a given value of $T/T_{c0}.$The lower panel
shows the correction at $T=0$ and a clear upturn of the conductivity which is
characteristic of the quantum regime. This would correspond to a decrease in
resistance ( negative magnetoresistance) as $\alpha$ is increased at $T=0$ and
a clearly visible non-monotonic behavior at low temperatures.}%
\label{2dalphaplot}%
\end{center}
\end{figure}

In Figs. \ref{2dTplot} and \ref{2dalphaplot} we display the plots for the
fluctuation conductivity when the vicinity of the pair-breaking quantum phase
transition is explored either by sweeping the temperature or the pair-breaking
parameter (see figure captions for more details). The behavior of the
fluctuation conductivity is qualitatively similar to the case of $d=1$.
However the critical divergences are weaker and the ratio of the fluctuation
correction to the normal state conductivity is expected to be lower as well.

\subsubsection{Bulk (d = 3)}

Although the pair-breaking quantum phase transition out of a superconducting
state in low dimensional disordered systems attracts more attention, we will
now consider such a transition in a three-dimensional bulk superconductor
mainly for the sake of comparison and completeness. One could imagine
superconductivity at low temperatures being destroyed by magnetic impurities,
for example. The discussion below parallels the preceding analysis for $d=1,2$.

The fluctuation conductivity in the classical regime (Eq. (\ref{ALsh>})) is
given by
\begin{align}
{\delta\sigma_{>,sh}^{AL}}(\alpha,T)  & =\frac{Te^{2}}{6\pi^{2}\sqrt{D}%
\sqrt{\alpha_{c}}}\int_{0}^{\infty}dx\frac{x^{4}}{\left(  \delta_{\alpha
}(\alpha,T)+x^{2}/2\right)  ^{3}}\nonumber\\
& =\ \frac{e^{2}}{4\pi\sqrt{2}\sqrt{D}}\frac{T}{\ \sqrt{\alpha-\alpha_{c}(T)}}%
\end{align}
and at $\alpha=\alpha_{c0}$ reduces to
\begin{equation}
{\delta\sigma_{>,sh}^{AL}}(\alpha_{c0},T)=\frac{\sqrt{3}{e^{2}}\sqrt
{\alpha_{c0}}}{\sqrt{D}4\pi^{2}}\label{noQCdiv3d}%
\end{equation}
which is independent of temperature. This is in stark contrast to a quantum
critical divergence found in the case of one and two dimensions on approaching
the quantum phase transition by coming down in temperature. It illustrates the
fact that fluctuations are stronger and consequently play a more crucial role
in reduced dimensions.

Same as for a nanowire and a thin film, note that the fluctuation conductivity
Eq. (\ref{ALsh_class}) near the classical transition at $T_{c0}$ in the
absence of any pair-breaking perturbation is analogous in form to that in the
classical regime and is given by
\begin{equation}
{\delta\sigma_{sh}^{AL}}(0,T)=\frac{e^{2}\ }{8\sqrt{2\pi}\sqrt{D}\ }%
\frac{T_{c0}}{\sqrt{T-T_{c0}}}%
\end{equation}
Although it is critical, the divergence is weaker by one power of
$(T-T_{c0})^{-1/2}$as compared to that in two dimensions, which in turn is
weaker than that in one dimension by the same power.

By using Eq. (\ref{ALsh<}) the fluctuation conductivity in the intermediate
regime is given by
\begin{align}
{\delta\sigma_{<,sh}^{AL}(\alpha,T)}  & {=}\frac{T^{2}e^{2}}{9\pi\sqrt
{D}\alpha_{c}^{3/2}}\int_{0}^{\infty}dx\frac{x^{4}}{\left(  \delta_{\alpha
}(\alpha,T)+x^{2}/2\right)  ^{4}}\ \nonumber\\
& =\ \frac{e^{2}}{36\sqrt{2}\sqrt{D}}\frac{T^{2}}{\ (\alpha-\alpha
_{c}(T))^{3/2}}%
\end{align}
It contains an additional power of $T/(\alpha-\alpha_{c}(T))$ as compared to
the classical regime, just as we found for $d=1,2$. \ Alternatively expressing
${\delta\sigma_{sh}^{AL}(\alpha,T)}$ in terms on the scaling function that we
introduced in Eq. (\ref{ALshSF}) we have
\begin{equation}
{\delta\sigma_{sh}^{AL}(\alpha,T)=\ \frac{8\sqrt{2}{e}^{2}\sqrt{T}}{3{\pi
}\sqrt{D}}}\ F(\eta)
\end{equation}
with
\begin{equation}
F(\eta)=\left\{
\begin{array}
[c]{cccc}%
\frac{3}{64}\eta^{-1/2} &  &  & \eta\ll1\\
\ \frac{\pi}{192}\ \eta^{-3/2} &  &  & \eta\gg1
\end{array}
\right.
\end{equation}
and the full form of $F(\eta)$ can be evaluated by doing the integrals in Eq.
(\ref{ScalFunc}) numerically.%

\begin{figure}
[ptb]
\begin{center}
\includegraphics[
height=3.2145in,
width=3.0475in
]%
{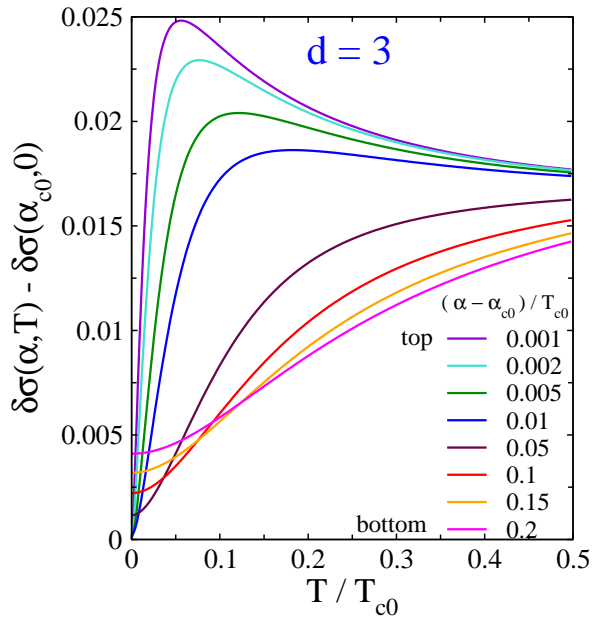}%
\caption{Temperature dependence of the fluctuation correction to the
conductivity (in units of $e^{2}/\sqrt{D}$) in the vicinity of a pair-breaking
quantum phase transition at $\alpha_{c0}=0.889T_{c0}$, for the case of a bulk
system ($d=3$). Plots are shown for different values of $(\alpha-\alpha
_{c0})/T_{c0}$. When the superconducting QCP is approached by lowering the
temperature at $\alpha=\alpha_{c0}$, there is no quantum critical divergence
of the conductivity in contrast to what is found for $d=1,2$ and in fact the
conductivity is temperature independent (see Eq. (\ref{noQCdiv3d}). Note that
we have plotted the difference $\delta\sigma(\alpha,T)-\delta\sigma
(\alpha_{c0},0)$ which is always positive given that $\delta\sigma(\alpha,0)$
is most negative at $\alpha=\alpha_{c0}$ (see Fig. \ref{3dalphaplot}). }%
\label{3dTplot}%
\end{center}
\end{figure}

The temperature independent quantum correction which dictates the behavior in
the quantum regime can be obtained by using Eq. (\ref{sigmazeroT}) to have
\begin{equation}
\delta\sigma_{0,cth}(\alpha)=-{\frac{{e^{2}}\sqrt{\alpha_{c0}}}{15{\pi}%
^{3}\sqrt{D}}}\int_{0}^{\infty}{dx}\frac{x^{6}}{h_{\alpha}(x)^{3}\ln
h_{\alpha}(x)}%
\end{equation}
where $h_{\alpha}(x)$ is defined by Eq. (\ref{h}). The integral can again be
regulated by considering the difference $\delta\sigma_{0,cth}(\alpha
)-\delta\sigma_{0,cth}(\alpha_{c0})$ as we did in the case of $d=2$, and can
then be subjected to numerical evaluation.

By following the same procedure and in one and two dimensions we get the
expansion
\begin{align}
& \delta\sigma_{0,cth}(\alpha)-\delta\sigma_{0,cth}(\alpha_{c0})\\
& =\frac{e^{2}\sqrt{\alpha_{c0}}}{\ \sqrt{D}}[a_{3}\text{ }\delta_{\alpha
}(\alpha,0)\ \nonumber\\
& +b_{3}\delta_{\alpha}(\alpha,0)^{2}+c_{3}\delta_{\alpha}(\alpha
,0)^{5/2}+...]\nonumber
\end{align}
with $a_{3}=0.023$ and $b_{3}=-0.061$ and $c_{3}=4\sqrt{2}/(15\pi^{2})$. Note
that the critical divergence shows up for the first time in the third
derivative to give the first non-analytic term in the expansion. By using the
first term in the expansion, one can make a rough estimate of the boundary
between the intermediate and the quantum regimes to be
\begin{equation}
\frac{T_{0}(\alpha)}{T_{c0}}\sim\left(  \frac{\alpha-\alpha_{c0}}{\alpha_{c0}%
}\right)  ^{5/4}\label{3dQIbound}%
\end{equation}
The quantum regime is thus expected to extend up to higher temperatures as
compared to the case of $d=1,2$. The fluctuation conductivity however has a
weaker critical divergence as compared to lower dimensions as was found above
also for the classical and intermediate regimes.

\begin{figure}
[ptb]
\begin{center}
\includegraphics[
height=4.1677in,
width=3.094in
]%
{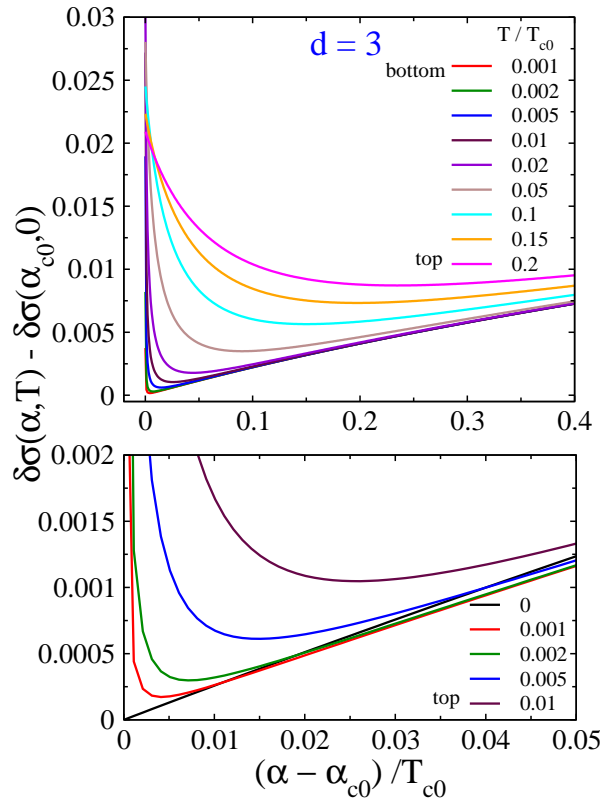}%
\caption{Fluctuation correction to the conductivity (in units of $e^{2}%
/\sqrt{D}$) in the vicinity of a pair-breaking quantum phase transition at
$\alpha_{c0}=0.889T_{c0}$ for $d=3$ (bulk superconductors). Plots show the
behavior of $\delta\sigma(\alpha,T)-\delta\sigma(\alpha_{c0},0)$ as one sweeps
$(\alpha-\alpha_{c0})/T_{c0}$ at a given value of $T/T_{c0}.$The lower panel
shows the correction at $T=0$ and a clear upturn of the conductivity which is
characteristic of the quantum regime. This would correspond to a decrease in
resistance ( negative magnetoresistance) as $\alpha$ is increased at $T=0$ and
a clearly visible non-monotonic behavior at low temperatures.}%
\label{3dalphaplot}%
\end{center}
\end{figure}

In Figs. \ref{3dTplot} and \ref{3dalphaplot} we display the plots for the
fluctuation conductivity when the vicinity of the pair-breaking quantum phase
transition is explored either by sweeping the temperature or the pair-breaking
parameter. The qualitative behavior is similar to that discussed for the case
of one and two dimensions, but the critical divergences are clearly much
weaker. The biggest difference in the behavior can be seen if the quantum
phase transition at $\alpha_{c0}$ is approached by coming down in temperature:
as shown above, there is absolutely no quantum critical divergence and the
correction is independent of temperature.%

\section{Related work, experiments and conclusion}

In our work we have studied the superconducting fluctuation corrections to the
normal state conductivity in the entire $\alpha$-$T\ $plane, where $\alpha$
parametrizes the strength of a pair-breaking perturbation, caused by the
presence of magnetic impurities or a magnetic field, for example. We have been
particularly interested in mapping out the fluctuation regimes in the
neighborhood of the pair-breaking quantum phase transition from
superconducting to normal state. Our entire analysis has been carried out
within the framework of temperature diagrammatic perturbation theory suitable
for dirty superconductors and disordered systems.

There have been two previous works which have used an approach different than
ours. Ramazashvili and Coleman\cite{RamazashviliC1997} have used an effective
action for the pairing field and evaluated the Aslamazov-Larkin correction to
the conductivity using a renormalization group analysis\cite{Millis1993}. They
have focussed on the quantum phase transition driven by magnetic impurities in
two and three dimensions for weak (BCS) as well as strong coupling
superconductors and derived the behavior of the conductivity when the quantum
critical point is approached by lowering the temperature. The temperature
dependence matches with what we obtain is two dimensions (see Eq.
(\ref{QCdivd2})) but in three dimensions our answers do not match. While they
have considered only the \textquotedblleft classical renormalization
region\textquotedblright, Mineev and Sigrist\cite{MineevS2001} have
investigated the entire region around the quantum phase transition, using an
analogous starting point based on the time-dependent Ginzburg-Landau equations
(justifying the use based on arguments by Herbut\cite{Herbut2000}). They
suggest pressure as a tuning parameter, consider weak coupling superconductors
in one, two and three dimensions, and obtain the conductivity corrections in
what they call, \textquotedblleft classical\textquotedblright\ and
\textquotedblleft quantum\textquotedblright\ regimes. Their classical regime
is identical to ours, while the answers they find in their quantum regime
correspond to our intermediate regime.

The above-mentioned approaches based on using the time-dependent
Ginzburg-Landau equations with a linear dissipative time derivative or the
corresponding effective action are quite powerful and analogous approaches
have proved extremely useful in the study of quantum critical
phenomena.\cite{Hertz1976,Millis1993,Sachdev} However, the Maki-Thompson and
the density-of-states corrections that require the electrons to be present in
the theory, are outside the scope of these approaches and a microscopic
calculation, as we have carried out here, becomes essential to analyze the
role of these corrections. Even the zero-temperature Aslamazov-Larkin
correction that we find is missed in the effective approaches. On the other
hand, our analysis is able to identify the regimes where the contributions
involving the interplay between the fluctuating Cooper pairs and the electrons
are subdominant, thereby validating the use of the afore mentioned effective
approaches in those regimes.

We want to now mention some works which also follow a microscopic approach,
but for physical configurations different than ours. Beloborodov and
Efetov\cite{BeloborodovE1999,BeloborodovEL2000} have proposed the negative
magnetoresistance observed in granular metals in a strong magnetic field and
low temperatures, to be originating from superconducting fluctuation
corrections to the conductivity. Galitski and Larkin\cite{GalitskiL2001a} have
considered two-dimensional superconductors in the presence of a perpendicular
magnetic field and again carried out a microscopic analysis of fluctuation
corrections taking into account all the diagrams; they too find zero
temperature negative magnetoresistance. As we have discussed in Ref.
[\onlinecite{LopatinSV2005}]%
, negative magnetoresistance is found also for a thin film in a parallel
instead of perpendicular magnetic field. We think it is quite a remarkable
fact that a negative magnetoresistance at zero temperature is a common feature
of all these theories: the Aslamazov-Larkin correction which is always
positive, the density-of states correction which is always negative and the
Maki-Thompson correction which has no prescribed sign, conspire in all three
theories to add up into a total negative correction. Although there might be
different reasons for getting a negative correction and a corresponding
negative magnetoresistance, it raises the question whether this a universal
feature of at least a certain class of disordered systems in the presence of a
magnetic field/pair-breaking perturbation, with fundamental reasons at its heart.

In our work we have considered the corrections to the conductivity coming into
effect due to the proximity of a superconducting state in the low temperature
phase diagram of disordered systems. It is important for our analysis that the
disordered conductors under consideration are assumed to be in the metallic
(as against insulating)\ conduction domain. The fact that quantum corrections
coming from weak localization effects\cite{GorkovLK1979} and electron-electron
interactions (the so-called Altshuler-Aronov\cite{AltshulerAL1980}
corrections) significantly modify the conductivity from its Drude-like
behavior, even in this domain, does not invalidate our analysis. However in
comparing our theory with experiments, these corrections should be
simultaneously taken into account. The weak localization of electron waves is
a result of an enhanced back-scattering originating from the interference
between forward and backward electron trajectories tracing the same path
during the course of multiple scattering events. It gives rise to a negative
quantum correction to the conductivity, given by the expression
\cite{GorkovLK1979}%
\begin{equation}
\delta\sigma_{WL}=-\frac{2e^{2}D}{\pi}\int\frac{d^{d}q}{(2\pi)^{d}}C(0,q)
\end{equation}
As is evident from the presence of a Cooperon propagator Eq. (\ref{Cooperon1}%
), the interference effects are strongly diminished by the presence of
time-reversal symmetry breaking perturbations, thereby yielding a decrease in
resistance (a negative magnetoresistance, yet again). On the other hand, the
Altshuler-Aronov correction is determined by the diagrams that include only
the Diffuson propagators and are not sensitive to the magnetic field or other
time-reversal symmetry breaking perturbations. Thus the inclusion of
additional quantum corrections will not affect the predicted negative sign of
the magnetoresistance (or the decrease of resistance with increasing
pair-breaking strength, in the general case) at low temperatures.

The experimental effort in superconducting thin films has so far been
motivated to a large extent by interest in the so-called
superconducting-insulator transition (SIT)\cite{HavilandLG1989}, driven by
tuning either the disorder (achieved by varying the film thickness) or a
perpendicular magnetic field. Questions raised by recent experiments about the
mechanism and interpretation of the transition, has revived interest also from
the theoretical side. Experiments on thin films, observing a suppression of
superconductivity in the presence of a perpendicular magnetic field at low
temperatures, have conventionally been interpreted within the framework of the
field-induced dual-SIT (acronym we will use to refer to a theory based on the
boson-vortex duality in a \textquotedblleft dirty boson\textquotedblright%
\ model\cite{Fisher1990}, although in the literature such a theory is
implicitly implied whenever the acronym SIT\ is used) mainly based on the
negative temperature derivative of the resistance above the critical field and
finite size scaling analysis of the data. Gantmakher et
al.\cite{GantmakherETZB2003} have made the case that a stringent analysis of
many of these data sets might point towards inadequacies in such an
interpretation. For instance, the phase interpreted as insulating could very
well be a metal with quantum corrections; and existence of scaling in a
limited region might not be sufficient. To address these points, they have
made measurements on NdCeCuO films and found that the microscopic theory based
on quantum corrections including the superconducting fluctuation corrections
as obtain by Galitski and Larkin\cite{GalitskiL2001a} could qualitatively
describe the main features of their experiment including the negative
magnetoresistance (see the discussion in preceding paragraphs). A\ possible
crossover between the two interpretations is also suggested. Subsequently,
Baturina et al.\cite{BaturinaIBSBS2004} have made measurements of
magnetic-field-dependent resistance of ultrathin superconducting TiN films
with different degrees of disorder and again concluded that the scaling
analysis previously regarded as the main evidence of field-induced dual-SIT
can in fact be observed also for transition from a superconductor to a normal
metal with quantum corrections.

The natural question to ask is, what happens when superconductivity at very
low temperatures is destroyed by applying not a perpendicular but instead a
parallel magnetic field for which the field-tuned dual-SIT\ scenario does not
apply? Although there is an interesting set of experiments studying the
first-order spin paramagnetic transition (e.g. Ref.
[\onlinecite{Adams2004}]%
), the case of relevance to us is that of a second order transition. Based on
their measurements on InO films with variable oxygen content, Gantmakher et
al.\cite{GantmakherGDST2000} concluded that the behavior in the parallel and
perpendicular field-tuned case is very similar. The non-monotonic
magnetoresistance they find in both cases, is interesting given that we
theoretically find a similar behavior in the case of a parallel magnetic
field, as presented in the previous section. Parendo et
al.\cite{ParendoHBG2004} have carried out an experiment on ultrathin bismuth
films to study the thickness-tuned SIT but in the presence of a parallel
magnetic field. In the immediate vicinity of the transition, what they find is
a negative magnetoresistance behavior. Based on the analysis of their data,
they argue that perhaps its origin could be found in the negative fluctuation
corrections to the conductivity that we find in our calculation. It does seem
plausible that the conductivity behavior near the superconducting transition
tuned by varying the thickness at a fixed value of parallel magnetic field
might be closely related to that tuned by varying the parallel magnetic field
at a fixed value of film thickness. However a definitive calculation catering
to the former case still remains to be done. As is evident from the expression
Eq. (\ref{alpha_film}), the pair-breaking parameter depends not only on the
magnetic field but also on the thickness of the film. If the only effect of
changing the thickness were to tune the pair-breaking strength, then the two
cases would in fact be identical. The complication arises because tuning the
thickness results also in tuning the disorder strength. Tuning the magnetic
field offers a way of isolating the pair-breaking effect. More recently, Aubin
et al.\cite{AubinMPBBDL2006} have done measurements of field-tuned SIT\ on
NbSi thin films and interpreted the data to conclude that in their experiment,
the case of perpendicular field is different from that of parallel, based on
the presence or absence of a kink in the temperature profile of the critical
field, respectively.

It is desirable to have a systematic experimental study aimed specifically at
exploring the physics of a pair-breaking quantum phase transition in
superconducting films. There have been few works using a parallel field, but
they have had a limited goal focussing on the SIT, as mentioned in the
preceding paragraph. Amorphous (non-granular) superconducting films that are
thinner than $\xi$, but not too thin; and are weakly disordered, with as low a
sheet resistance as possible; would be necessary to assure that the quantum
corrections are small enough. To begin with, it will be useful to observe the
finite temperature classical transition and verify the predictions of the
fluctuation conductivity in its vicinity. By slowly increasing the
pair-breaking strength and lowering the temperature, one could approach the
quantum phase transition. Having identified the right films, measurements of
the temperature and pair-breaking parameter (tuned by a parallel field, for
example)\ dependence of the conductivity, would afford an exciting possibility
of discovering different regimes in the vicinity of the pair-breaking quantum
phase transition. The increase in normal state resistance due to the presence
of superconducting fluctuations that we find, is in stark contrast to the
intuitive expectation and is a purely quantum effect. A clear experimental
signature of such a characteristically quantum behavior in the
\textit{quantum} regime, changing over into an increase in conductivity in the
\textit{classical} regime, would be an important step forward in the study of
quantum phase transitions and low temperature superconductivity. The
manifestation to be expected in the plots for conductivity as a function of
temperature and pair-breaking strength can be found in the previous section.
The next revealing experiment would be to measure the evolution of the
conductivity behavior with the change of angle made by the magnetic field with
the film, ranging from parallel to perpendicular case.

Though most experiments on thin films have focussed on perpendicular magnetic
field and disorder tuned transitions, recently Parker et
al.\cite{ParkerRKX2006} have performed measurements on homogeneously
disordered ultrathin a-Pb films to study the magnetic impurity tuned
transition in addition. They have compared the conductivity behavior near
quantum phase transitions tuned by all three mechanisms. They have concluded
that the disorder tuned and the magnetic impurity tuned cases show similar
behavior which seems to be consistent with a fermionic nature of the
transition to a weakly insulating normal state. Their experiment is of
relevance to us since they have successfully traced the transition line and
shown that it satisfies the Abrikosov-Gorkov suppression of $T_{c}$ given by
Eq. (\ref{Tc_vs_H}). However the dependence of the conductivity on the
magnetic impurity concentration has not been presented and the temperature
dependence has been shown only for one concentration value that is above the
critical concentration corresponding to $\alpha_{c0}$. It will be very
interesting to carry out a detailed comparison of this experiment with our theory.

Systematic experiments exploring the pair-breaking quantum phase transition
tuned by magnetic impurities in three dimensional superconductors would
provide a good check of our theory, given that in this case we are above the
upper critical dimension. For example, verifying the lack of quantum critical
divergence as the critical impurity concentration is approached by coming down
in temperature and finding the temperature independent behavior instead, would
provide a contrast to the critical divergence expected in the case of one and
two dimensional superconductors. Finding a weak non-monotonic behavior and the
presence of a quantum regime would be interesting in its own right. In
addition, the analysis of the transition in a relatively simple material would
provide useful insights in interpreting the behavior near the superconducting
quantum phase transitions in more complex materials.

As far as superconducting quantum phase transition and quantum corrections are
concerned, a lot of theoretical as well as experimental work on disordered
thin films has been carried out over the period of last two decades.
Relatively less work on similar lines has been done on superconducting wires.
However the technological advance allowing for the fabrication of
superconducting nanowires that are uniform and ultrathin ($<$ $10$ nm) (see
e.g. Ref.
[\onlinecite{BezryadinLT2000}]%
) has invigorated the field and opened up new possibilities. By coating carbon
nanotubes or DNA\ molecules suspended over a trench, with a superconducting
alloy such as MoGe or Nb, one essentially obtains superconducting wires
connected on two sides to thin film electrodes of the same material. So far
most experiments done on this set-up have focussed on unraveling the physics
of phase-slip fluctuations that result in a non-zero resistance below the
superconducting transition. In one recent experiment, Rogachev, Bollinger and
Bezryadin\cite{RogachevBB2005} have looked at the effect of magnetic fields.
They were able to explain the suppression of the transition temperature in
terms of the pair-breaking theory (Eq.~(\ref{transline})) if the Zeeman
pair-breaking effect in the presence of spin-orbit coupling was taken into
account in addition to the orbital effect of the magnetic field applied
perpendicular to the wire (see Eq.~(\ref{alphawireperp})). Although the
fluctuation effects above the transition were not explored until now, we
believe that the experimental set-up is appropriately geared to be able to
test the predictions of our theory. Systematic study of low temperature
fluctuations both below and above the transition would be essential not only
to understand the nature of mesoscopic superconductivity but also to be able
to successfully control superconducting electronic circuits involving
ultra-narrow wires.

The parallel-magnetic field tuned quantum phase transition in doubly-connected
superconducting cylinders that we discuss in Sec. II --with the pair-breaking
parameter given by Eq. (\ref{alpha_cyl})-- has been motivated by its
experimental realization by Liu et al.\cite{LiuZRRCW2001}. They clearly seem
to have observed the enhancement of the conductivity resulting from the
positive fluctuation contribution in the classical regime. But so far there
has been no experimental evidence of the corrections expected in the
intermediate and quantum regime. Further experiments in this direction would
be very interesting.

In conclusion, we have evaluated the fluctuation corrections to the electrical
conductivity in the vicinity of the pair-breaking quantum phase transition
using diagrammatic perturbation theory in disordered systems by correctly
incorporating the quantum fluctuations within the formalism. Amongst the three
distinct superconducting fluctuation regimes that we find, the quantum regime
is the one is which the contributions to the conductivity coming from the
interaction between the fluctuating pairs and normal electrons are important,
while the behavior in the classical (or quantum critical) and the intermediate
regime is dominated by direct transport via the fluctuating pairs. Our theory
thus seems to indicate that the effective bosonic action formalism should be
applicable outside of the quantum regime. It should be noted that even at zero
temperature, our theory is expected to be valid only outside the quantum
Ginzburg region for systems below or at the upper critical dimension. One
important open problem is to understand in what way the zero temperature
conductivity above the transition connects with that below the transition.

Within our microscopic theory, we have been able to extract the finite
temperature crossovers to be expected near the superconducting quantum
critical point. We have used the electrical conductivity as a means of probing
the effects of superconducting fluctuations, however a similar analysis could
also be carried out for the diamagnetism and other thermodynamic quantities.
Extension to the case of anisotropic superconductors such as high temperature
superconductors would also be interesting. In d-wave superconductors, disorder
acts as a pair-breaker and the behavior near the corresponding pair-breaking
quantum phase transition is likely to have some resemblance to our findings.

We believe that pair-breaking quantum phase transitions form an important
class of quantum phase transitions that not only allow for a systematic,
well-controlled experimental exploration but also provide the possibility of a
thorough and comprehensive theoretical analysis. We hope that our work has
amply demonstrated the latter and will in turn motivate the former. In the
end, we expect that the microscopic theory of the superconducting quantum
critical point and the corresponding experimental analysis would serve as a
useful prototype for understanding quantum phase transitions also in other
classes of correlated systems.

\textit{Acknowledgments} One of the authors (NS) would like to gratefully
acknowledge Alexey Bezryadin, Allen Goldman and Kevin Parendo for discussions
regarding experiments and the grant NSF DMR 0605813 for financial support
while part of the work was carried out.

\bibliographystyle{apsrev}
\bibliography{fluctQPT}

\end{document}